\documentstyle[12pt]{article}

\newcommand{\nc}{\newcommand}
\newcommand{\rnc}{\renewcommand}

\makeatletter
\def\section{\@startsection {section}{1}{\z@}{-3.5ex plus -1ex minus
 -.2ex}{2.3ex plus .2ex}{\large\bf}}
\def\subsection{\@startsection{subsection}{2}{\z@}{-3.25ex plus -1ex minus
 -.2ex}{1.5ex plus .2ex}{\normalsize\bf}}
\makeatother

\makeatletter
\@addtoreset{equation}{section}
\rnc{\theequation}{\thesection.\arabic{equation}}
\makeatother

\nc{\ignore}[1]{}

\nc{\be}{\begin{equation}}
\nc{\ee}{\end{equation}}
\nc{\bea}{\begin{eqnarray}}
\nc{\eea}{\end{eqnarray}}
\nc{\ol}[1]{\overline{#1}}

\rnc{\a}{\alpha}
\rnc{\b}{\beta}
\rnc{\d}{\delta}
\nc{\ep}{\epsilon}
\nc{\e}{\eta}
\nc{\eb}{\bar{\eta}}
\nc{\f}{\phi}
\nc{\fb}{\bar{\phi}}
\nc{\vf}{\varphi}
\nc{\p}{\psi}
\rnc{\pb}{\bar{\psi}}
\rnc{\c}{\chi}
\nc{\cb}{\bar{\c}}
\nc{\la}{\lambda}
\nc{\m}{\mu}
\nc{\n}{\nu}
\rnc{\o}{\omega}
\rnc{\t}{\theta}
\nc{\tb}{\bar{\theta}}
\nc{\eps}{\epsilon}
\nc{\Om}{\Omega}
\nc{\ga}{\gamma}
\rnc{\pd}{\partial}

\rnc{\S}{\Sigma}
\nc{\Sa}{\S\times\{0\}}
\nc{\Sb}{\S\times\{1\}}
\nc{\SI}{\S\times I}
\nc{\SS}{\S\times S^{1}}
\nc{\Sg}{\S_{g}}
\nc{\M}{{\cal M}}
\nc{\MF}{\M_{{\cal F}}}

\nc{\trac}[2]{{\textstyle\frac{#1}{#2}}}

\nc{\ex}[1]{{\rm e}^{\,\textstyle#1}}

\def\slash#1{\setbox0=\hbox{$#1$}#1\hskip-\wd0\hbox to\wd0{\hss\sl/\/\hss}}

\nc{\mat}[4]{\left(\begin{array}{cc}#1&#2\\#3&#4\end{array}\right)}

\def\tr{\mathop{\rm tr}\nolimits}
\def\Tr{\mathop{\rm Tr}\nolimits}
\nc{\ra}{\rightarrow}
\nc{\ot}{\otimes}
\rnc{\ss}{\subset}
\nc{\ad}{{\rm ad}}
\nc{\Ad}{{\rm Ad}}
\rnc{\lg}{{\bf g}}
\nc{\lt}{{\bf t}}
\nc{\lk}{{\bf k}}
\nc{\lf}{{\bf f}}
\nc{\lh}{{\bf h}}
\nc{\bft}{{\bf t}}
\nc{\bfk}{{\bf k}}
\nc{\bfg}{{\bf g}}
\nc{\del}{\partial}
\nc{\dbar}{\bar{\del}}
\nc{\dz}{\del_{z}}
\nc{\zb}{\bar{z}}
\nc{\dzb}{\del_{\bar{z}}}
\nc{\az}{A_{z}}
\nc{\azb}{A_{\bar{z}}}
\nc{\bz}{B_{z}}
\nc{\bzb}{B_{\bar{z}}}
\nc{\ba}{{\bf A}}
\nc{\bb}{{\bf B}}
\nc{\g}{g^{-1}}
\nc{\dw}{\Delta_{W}}% the weyl determinant
\nc{\Det}{{\rm Det}\,} %functional \det
\nc{\ddw}{\det\dw}
\nc{\Ddw}{\Det\dw}
\nc{\bG}{{\bf G}}
\nc{\bT}{{\bf T}}
\nc{\bF}{{\bf F}}
\nc{\bH}{{\bf H}}
\nc{\unA}{\underline{A}}
\nc{\C}{{\cal A}/{\cal G}}
\nc{\A}[1]{{\cal A}^{#1}/{\cal G}^{#1}}
\nc{\dx}{\dot{x}}
\rnc{\O}[2]{\Omega^{#1}({#2},\lg)}
\nc{\wif}{Weyl integral formula}
\nc{\CS}{Chern-Simons}

\def\sAA{{\rm A\kern-0.85em A}} % simple version
\def\tAA{{\mathchoice
  {\sAA}
  {\sAA}
  {\rm A\kern-0.60em A}
  {\rm A\kern-0.50em A} }}
\def\sBB{{\rm I\kern-.17em{}B}}
\def\BB{{\mathchoice
  {\sBB}
  {\sBB}
  {\rm I\kern-.13em{}B}
  {\rm I\kern-.13em{}B} }}
\def\sCC{{\kern 0.27em\vrule height1.45ex width0.03em depth0em
	  \kern-0.30em\rm C}}
\def\CC{{\mathchoice
  {\sCC}
  {\sCC}
  {\kern 0.225em \vrule height1.05ex width0.025em depth0em \kern-0.25em \rm C}
  {\kern 0.180em \vrule height0.78ex width0.02em depth0em \kern-0.2em \rm C}
	}}
\def\tCC{{\ooalign{C\crcr\kern0.27em\vrule height1.45ex width0.03em
depth0em\crcr}}}
\def\sDD{{\rm I\kern-.16em{}D}}
\def\DD{{\mathchoice
  {\sDD}
  {\sDD}
  {\rm I\kern-.13em{}D}
  {\rm I\kern-.13em{}D} }}
\def\sEE{{\rm I\kern-.17em{}E}}
\def\EE{{\mathchoice
  {\sEE}
  {\sEE}
  {\rm I\kern-.13em{}E}
  {\rm I\kern-.13em{}E} }}
\def\sFF{{\rm I\kern-.16em{}F}}
\def\FF{{\mathchoice
  {\sFF}
  {\sFF}
  {\rm I\kern-.13em{}F}
  {\rm I\kern-.13em{}F} }}
\def\sGG{{\kern 0.27em \vrule height1.45ex width0.03em depth0em
	  \kern-0.30em \rm G}}
\def\GG{{\mathchoice
  {\sGG}
  {\sGG}
  {\kern 0.225em \vrule height1.05ex width0.025em depth0em \kern-0.25em \rm G}
  {\kern 0.180em \vrule height0.78ex width0.020em depth0em \kern-0.20em \rm G}
	}}
\def\sHH{{\rm I\kern-.16em{}H}}
\def\HH{{\mathchoice
  {\sHH}
  {\sHH}
  {\rm I\kern-.13em{}H}
  {\rm I\kern-.13em{}H} }}
\def\sII{{\rm I\kern-.16em{}I}}
\def\II{{\mathchoice
  {\sII}
  {\sII}
  {\rm I\kern-.12em{}I}
  {\rm I\kern-.10em{}I} }}
\def\sJJ{{\kern0.17em\vrule height1.5ex width 0.03em depth0em
	  \kern-.20em\rm J}}
\def\JJ{{\mathchoice
  {\sJJ}
  {\sJJ}
  {\kern0.150em\vrule height1.05ex width 0.025em depth0em\kern-.175em\rm J}
  {\kern0.135em\vrule height0.78ex width 0.020em depth0em\kern-.155em\rm J} }}
\def\sKK{{\rm I\kern-.16em{}K}}
\def\KK{{\mathchoice
  {\sKK}
  {\sKK}
  {\rm I\kern-.12em{}K}
  {\rm I\kern-.10em{}K} }}
\def\sLL{{\rm I\kern-.16em{}L}}
\def\LL{{\mathchoice
  {\sLL}
  {\sLL}
  {\rm I\kern-.12em{}L}
  {\rm I\kern-.10em{}L} }}
\def\sMM{{\rm I\kern-.16em{}M}}
\def\MM{{\mathchoice
  {\sMM}
  {\sMM}
  {\rm I\kern-.12em{}M}
  {\rm I\kern-.10em{}M} }}
\def\sNN{{\rm I\kern-.16em{}N}}
\def\NN{{\mathchoice
  {\sNN}
  {\sNN}
  {\rm I\kern-.12em{}N}
  {\rm I\kern-.10em{}N} }}
\def\sOO{{\kern 0.27em \vrule height1.50ex width0.03em depth0em
					\kern-0.30em \rm O}}
\def\OO{{\mathchoice
  {\sOO}
  {\sOO}
  {\kern 0.225em \vrule height1.05ex width0.025em depth0em \kern-0.25em \rm O}
  {\kern 0.180em \vrule height0.78ex width0.020em depth0em \kern-0.20em \rm O}
	}}
\def\sPP{{\rm I\kern-.16em{}P}}
\def\PP{{\mathchoice
  {\sPP}
  {\sPP}
  {\rm I\kern-.12em{}P}
  {\rm I\kern-.10em{}P} }}
\def\sQQ{{\kern 0.27em \vrule height1.45ex width0.03em depth0em
	  \kern-0.30em \rm Q}}
\def\QQ{{\mathchoice
	{\sQQ}
	{\sQQ}
  {\kern 0.225em \vrule height1.05ex width0.025em depth0em \kern-0.25em \rm Q}
  {\kern 0.180em \vrule height0.78ex width0.020em depth0em \kern-0.20em \rm Q}
	}}
\def\sRR{{\rm I\kern-0.16em{}R}}
\def\RR{{\mathchoice
  {\sRR}
  {\sRR}
  {\rm I\kern-0.12em{}R}
  {\rm I\kern-0.10em{}R} }}
\def\sSS{{\rm S\kern-.45em{}S}}
\def\sTT{{\rm T\kern-.60em{}T}}
\def\TT{{\mathchoice
  {\sTT}
  {\sTT}
  {\rm T\kern-.45em{}T}
  {\rm T\kern-.38em{}T} }}
\def\sUU{{\rm U\kern-.60em{}U}}
\def\UU{{\mathchoice
  {\sUU}
  {\sUU}
  {\rm U\kern-.46em{}U}
  {\rm U\kern-.38em{}U} }}
\def\sVV{{\rm V\kern-.62em{}V}}
\def\VV{{\mathchoice
  {\sVV}
  {\sVV}
  {\rm V\kern-.46em{}V}
  {\rm V\kern-.38em{}V} }}
\def\sWW{{\rm W\kern-.92em{}W}}
\def\WW{{\mathchoice
  {\sWW}
  {\sWW}
  {\rm W\kern-.80em{}W}
  {\rm W\kern-.67em{}W} }}
\def\sXX{{\rm X\kern-.58em{}X}}
\def\XX{{\mathchoice
  {\sXX}
  {\sXX}
  {\rm X\kern-.45em{}X}
  {\rm X\kern-.38em{}X} }}
\def\sYY{{\rm Y\kern-.58em{}Y}}
\def\YY{{\mathchoice
  {\sYY}
  {\sYY}
  {\rm Y\kern-.45em{}Y}
  {\rm Y\kern-.40em{}Y} }}
\def\sZZ{{\rm Z\kern-0.32em{}Z}}
\def\ZZ{{\mathchoice
  {\sZZ}
  {\sZZ}
  {\rm Z\kern-0.30em{}Z}
  {\rm Z\kern-0.25em{}Z} }}

\begin{document}
\global\parskip=4pt

%%%%%%%%% title page %%%%%%%%%%%%%%%%%%%%%%%%%%%%%%%%%%%%%%%%

\begin{titlepage}
\newlength{\titlehead}
\settowidth{\titlehead}{NIKHEF-H/91}
\begin{flushright}
%Last Update\\
%\today
IC/95/339\\
hep-th/9511038
\end{flushright}

\vskip .25in

\begin{center}
\vskip .25in
{\bf New Results in Topological Field Theory and Abelian Gauge Theory} \\
\end{center}

\vskip .25in

\begin{center}
\vskip .25in
{\bf George Thompson\footnote{thompson@ictp.trieste.it} } \\
I.C.T.P. \\
P.O. Box 586 \\
Trieste 34014\\
Italy \\
\end{center}

\begin{abstract}
These are the lecture notes of a set of lectures delivered at the 1995
Trieste summer school in June. I review some recent work on duality in
four dimensional Maxwell theory on arbitrary four manifolds, as well as
a new set of topological invariants known as the Seiberg-Witten
invariants. Much of the necessary background material is given,
including a crash course in topological field theory, cohomology of
manifolds, topological gauge theory and the rudiments of four manifold
theory.  My main hope is to wet the readers appetite, so that he or she
will wish to read the original works and perhaps to enter this field.
\end{abstract}

\end{titlepage}

%%%%%%%%% end of title page %%%%%%%%%%%%%%%%%%%%%%%%%%%%%%

\tableofcontents
\setcounter{footnote}{0}

\section{Introduction}
String theory is claimed to be the theory of `everything'. In principle,
within the context of string theory, one can calculate any process one
can dream of and also some processes, involving zero mass black holes
which were previously quite unimaginable. This makes the study of string
theory both fascinating and, in general, difficult.
Topological field
theory, on the other hand, is the theory of `nothing'. This means that
one does not calculate any physical transitions at all (at least not
directly). The reason for this being that there are no dynamical degrees
of freedom in these theories. This makes the
theory, in some instances, more tractable but, perhaps surprisingly,
does not diminish its inherent interest.

Why is a theory of `nothing' interesting? One aspect of the answer to this
question is that, even though
there are no physical degrees of freedom, these are fully interacting, and
at first sight, very complicated field theories. It comes as a surprise that
one can solve these models exactly. Nevertheless, this is so in many cases.
Solvable gauge field theories are few and far between yet Yang-Mills theory
and the $G/G$ coset models on a general Riemann surface and  Chern-Simons
theory and various relatives on arbitrary three manifolds are examples
\cite{rusakov,dfine,wit2ddon,btym,btrev,bjt}. We
can hope in this way to gain a better understanding of what a non-
perturbative solution to a field theory could be like.

What topological field theories calculate are invariants. That is
numbers that are robust: they are independent of couplings or of any
dynamics. These numbers are usually
the dimension of some space or an Euler character. These numbers may be
of interest to physicists or mathematicians or, indeed, to both. An example
of interest to both parties is Chern-Simons theory \cite{Witcs}. The partition
function of Chern-Simons theory
on a manifold of the form $\Sigma \times S^{1}$, where $\Sigma$ is a
Riemann surface, calculates the dimension of the space of conformal
blocks of the $G$ WZW model. For physicists this is the dimension of the
Hilbert space of the theory while for mathematicians this is the
dimension of the space of sections of the determinant line over the
space of flat connections on $\Sigma$ (somewhat of a mouthful). One can
calculate this directly in the
conformal theory, as E. Verlinde \cite{ver} did, but one only needs
familiar gauge
theory techniques from the Chern-Simons point of view and one bypasses
completely the conformal field theory technology.

While topological field theories are interesting in their own right
some of the interest in them is also due to their, rather direct,
relationship with
conventional ``physical'' theories. For example, one can
calculate Yukawa couplings in ($N=2$
supersymmetric-) string theory with the standard sigma model or with
either of the two possible topological theories\footnote{The calculation
of Yukawa couplings in string theory was
covered in Brian Greene's lectures. The relationship with the
topological theory comes about as one is restricting ones attention to
chiral primaries. On this restricted field set the supersymmetry
charges act like BRST operators.}. The topological field
theories can also act as ``easier'' testing grounds for ideas in
physical theories. An example of this is the idea of duality in
supersymmetric Yang-Mills theory in four dimensions. Suppose there is a
correspondence between the weak (strong) coupling of one theory and the
strong (weak) coupling of its dual theory. If one of these field theories
has a topological field theory hidden within it, then so does the other as
the topological field theory will exist in both phases of the original
theory. Now the form of the topological field theory may well be
different in the dual model. The equivalence of the two descriptions of
the topological field theories is then a necessary condition for the
duality of the starting models. In practice it may be easier to check for
duality at the level of the topological field theories.

This is the
situation studied by Vafa and Witten \cite{vafwit}. The twisted $N=4$ super
Yang-Mills theory on a four manifold calculates the Euler character of
the moduli space of instantons. Fortuitously, the mathematicians have
calculated these for certain compact K\"{a}hler manifolds as well as for
ALE spaces. Vafa and Witten were able to confirm, using the results of
the mathematicians, that indeed the partition function of the $N=4$
$SU(2)$ gauge theory transforms in strong coupling to the $N=4$ $SO(3)$
gauge theory at weak coupling. This provides a direct test of the
duality hypothesis.

Another example where one can use the topological theory (counter
historically) to test results
in a physical theory is provided by the candidate exact solution of
$N=2$ $SU(2)$ super Yang-Mills theory of Seiberg and Witten \cite{sw}.
The $N=2$
theory is related to some rather deep mathematics of four manifolds.
As explained by L. Alvarez-Gaume in his lectures, when one is at weak
coupling dominant contributions come from instantons. The bulk of the
mathematical analysis is to come to grips with the moduli space of
instantons. Rather than working at weak coupling $u \rightarrow \infty$
one can pass to the points $|u|=1$ where the physics is given in terms
of a massless monopole and the magnetic photon. This system is Abelian
and easily analysed, and should allow one, if the picture is correct, to
reproduce the Donoldson Polynomials. Witten \cite{witsw} shows in fact
that the theory at this point in the $u$ plane gives non-trivial
invariants of four manifolds (it is still a conjecture that one
reproduces Donaldson theory in general).

The list of contents spells out the order of things. However, I would
offer the following advice. Firstly, if you have no familiarity with
topological field theory you ought to skip to the example \ref{ecex} to
see how things work and then go back to the start. Secondly, the section
called a mathematical digression should be skimmed by those who are
familiar with cohomology, homology and their interrelation to check my
conventions. For those who are not fluent in these matters, I recommend
that they keep a copy of the report by Eguchi, Gilkey and Hanson
\cite{egh} close at hand.

These notes are very similar to sets of lectures presented earlier this
year in Trieste. The first was by Braam \cite{braam} and the second by
Dijkgraaf \cite{dijk}. The
reason for the overlap is easy to explain, our sources are almost
identical. The papers by Taubes \cite{taubes1}, Verlinde
\cite{versdual}, Witten \cite{witsdual,witsw} and others are very
clear and
hardly need elucidation -as for background the books by Freed and
Uhlenbeck \cite{fu} and especially that by Donaldson and Kronheimer
\cite{dk} are excellent
for the elaboration of many of the notions presented in the cited
publications. There are, however, also some differences in presentation.
These stem from the different audiences that we were addressing and from
the fact that I am not an expert in the field of four manifolds (whereas
the Dutch gentlemen cited are).

The papers I am describing in these lectures are really very nice and
one should also keep a copy of those nearby.

\vskip 1cm

\noindent\underline{Note Added}

Before, while and after delivering these lectures many related papers
appeared on the xxx archives. My aim is not to review all the literature
on the subject of Seiberg-Witten invariants; I apologise to those authors
that I do not mention. However, two reviews which complement these notes and
give the mathematics side of the story are \cite{ak,marc}.

\section{Topological Field Theories}
There are now a number of reviews on the general subject of topological
field theory that the reader may wish to consult
\cite{bbrt,blau,thompson,cmr}. However, the type of topological field
theories that will be of interest to us
here are easy to construct. We will want a field theory that devolves to
some moduli space, that is to say, that describes the space of solutions
to a set of equations. Suppose that we have a set of fields, $\{ \Phi_{i}
\}$ and the equations of interest are
\be
s^{a}(\Phi_{i})=0 \, , \label{equ}
\ee
we denote the space of solutions by ${\cal M}(\Phi)$. A
typical example is the space of flat connections on a manifold $X$. In
this case the fields $\Phi_{i}$ are only the gauge field $A$ and the
equations $s^{a}$ are
\be
s= F_{A}= dA + A^{2} = 0 \, . \label{flat}
\ee
Usually one wishes to also factor out the action of the gauge group and
this would correspond to yet more conditions on $A$ (a choice of gauge).
Another choice of interest is the space of self dual connections over a
4-manifold, the context in which Witten first introduced topological
field theories of this kind and to which we will pay some attention
later in these notes.

An important ingredient in the construction of a topological field
theory is the topological symmetry. We denote generator of the symmetry
by $Q$. Its action on the fields $\{ \Phi_{i} \}$ is to give a new set
of fields $\{ \Psi_{i} \}$ which are in all ways identical to the
original set except that the new fields have opposite
Grassmann character. If one starts with a scalar field then its
`superpartner' is also of spin zero, but it is an anticommuting field.
If, on the other hand, the field $\Phi$ is a spinor field
(anticommuting) then its partner $\Psi$ will be a commuting spinor
field, and so on.
This property of the set of fields $\{ \Psi_{i} \}$ makes them more like
Fadeev-Popov ghost fields and this is insinuated when one says that the
operator $Q$ is a BRST operator. So we have
\be
Q\Phi_{i} = \Psi_{i} .
\ee

In order to be able to impose the equations (\ref{equ}) we also need to
introduce a set of Grassmann odd fields $\{ \bar{\Psi}_{a} \}$ and
Grassmann even fields $\{ B_{a}  \}$ with transformation properties
\be
Q\ol{\Psi}_{a} = B_{a} .
\ee

Suppose that the theory we are interested in is defined on some manifold
(or more generally some space) $X$. In order to fully define the theory
we may need also to use a metric on $X$, or some other coupling
constants; denote these collectively by $t_{i}$. The action of interest
is then, schematically,
\bea
S & = &\int_{X} \{ Q, \left ( t_{0 }\ol{\Psi}_{a}s^{a}(\Phi_{i}) +
\sum_{i=1} t_{i} V^{i} \right) \} \nonumber \\
  & = & \int_{X} \left( t_{0} B_{a}s^{a}(\Phi) - t_{0} \ol{\Psi}_{a} \frac{ \d
s^{a}(\Phi)}{\d \Phi_{i}} \Psi_{i} + \dots \, \right)  . \label{topact}
\eea
The associated path integral is
\be
Z = \int DY \exp{(iS(Y))}
\ee
where $Y$ denotes all the fields.

Notice that if the multiplier fields $B_{a}$ appear as in (\ref{topact})
then integrating over them yield a delta function constraint on
$s^{a}=0$. Hence the partition function will devolve to a (finite
dimensional) integral over the moduli space. The integration over all of
the fields $\{ \overline{\Psi}^{a} \}$ will yield some function $\mu (
\Phi)$, so that
\be
Z= \int_{ {\cal M} } \mu (\Phi) ,
\ee
and $\mu(\Phi)$ may be interpreted as some measure on ${\cal M}$.

\noindent{\bf Remark:} It is sometimes possible to write down a
topological action without the need of an auxiliary set of fields. Two
dimensional Donaldson theory is an example of this \cite{wit2ddon,thompson}.

%\vspace{.8cm}

\subsection{The Moduli Space and its Co-Tangent Space}

We have already seen that the integral over $B_{a}$ imposes the
constraint that $s^{a}(\Phi)=0$, which in turn tells us that the path
integral over $\Phi$ finally becomes an integral over the moduli space.
Actually, because of the extra fields involved $\{ \Psi \}$ and $\{
\overline{ \Psi} \}$, one does not obtain, directly, just an integral
over the moduli space. In order to understand what space we are really
integrating over consider the integration over $\overline{\Psi}$, this
yields a delta function constraint on
\be
\frac{\d s^{a}(\Phi)}{\d \Phi_{i}} \Psi_{i} =0 . \label{tangent}
\ee
The fields $\Phi$ that appear in this formula should be understood to be
points in ${\cal M}$.

What (\ref{tangent}) tells us is that the $\Psi$ are properly thought of
as one-forms on the moduli space. To see this we note that one way to
determine a tangent vector $\d \Phi$ to ${\cal M}$ at a point $\Phi \in
{\cal M}$ is to compare to first order with a nearby point nearby point
$(\Phi  + \d \Phi) \in {\cal
M}$. As $s^{a}(\Phi + \d \Phi)=0$ we obtain the first oder equation for
$\d \Phi$,
\be
\frac{\d s^{a}(\Phi)}{\d \Phi_{i}} \d\Phi_{i} =0.
\ee
At first sight, on comparing this equation with (\ref{tangent}), we
would conclude that the $\Psi$ span the tangent space of ${\cal M}$,
however, that is not quite right. Rather, as the $\Psi$ are Grassmann
odd fields (relative to the $\Phi$) they are correctly thought of as
living in the cotangent bundle of $\cal M$, that is- they are one forms
on ${\cal M}$.

Interpreting the $\Psi$ as one-forms has the extra spin-off that now
there is a natural interpretation of the BRST operator $Q$ as well. The
formula
\be
Q\Phi_{i} = \Psi_{i}
\ee
tells us that acting on any functional $F[\Phi]$
\be
QF[\Phi] = \frac{\d F[\Phi]}{\d \Phi_{i}}\Psi_{i}. \label{qderiv}
\ee
On a finite dimensional manifold $X$ with local coordinates $x^{\mu}$ and
one forms $dx^{\mu}$ the exterior derivative $d$ acting on a function
$f(x)$ yields
\be
df(x)= \frac{df(x)}{dx^{\mu}}dx^{\mu}. \label{deriv}
\ee
A comparison of (\ref{qderiv}) and (\ref{deriv}) establishes that the
geometrical interpretation of $Q$ is as the exterior derivative.

This geometrical picture of a topological field theory and the
ingredients that go into making it is, in  essence, always the same.
Variations on the theme arise, when one must take some extra symmetry
into account, such as gauge invariance. We will come across examples of
this when considering topological field theories that model the moduli
space of instantons.

%\vspace{.8cm}

\subsection{Metric and Coupling Constant Independence}

In order to write down the equations of interest or an action one may need to
introduce extraneous parameters. For example, even if one can formulate
the flatness equation without recourse to a metric, a metric is,
nevertheless,  required in
order to gauge fix the gauge field. Another example is the instanton
equation which requires a metric from the word go. Under suitable
conditions one will be able to establish that nothing depends on these
choices. When that is so- the name of the game is to choose the parameters
to make life as simple as possible.

The variation of the partition function with respect to any of the
parameters $t_{i}$ (including the metric) is
\bea
\frac{\partial Z}{\partial t_{j}} &=& \int DY \exp{(iS(Y))} .
i \frac{\partial S(Y)}{\partial t_{j}} \nonumber \\
 &=& \int DY \exp{(iS(Y))} . i \{ Q , V(Y) \} ,
\eea
for some $V$. The right hand side vanishes by a Ward identity. Consider
the obvious equality
\be
\int DY \exp{(iS(Y))} V(Y) = \int D(Y + QY) \exp{iS(Y+QY)} V(Y+QY) .
\ee
This leads to non-trivial information if both the measure $DX$ and the
action are $Q$ invariant. In the topological field theory the action is
$Q$ exact, its Q variation will be $Q^{2}$ acting on something, but
$Q^{2}=0$, so $S(Y+QY)=S(Y)$. Presume the measure is also invariant (one
can check that at a formal level this is the case), so
that $D(Y+QY) = DY$. We can now conclude that
\bea
\int DY \exp{(iS(Y))} V(Y) &=& \int DY \exp{(iS(Y))} V(Y + QY) \nonumber
\\
          &=& \int DY \exp{(iS(Y))} \left( V(Y) + \{ Q , V(Y) \} \right)
\eea
which implies that
\be
\int DY \exp{(iS(Y))} \{ Q , V(Y) \} =0 .
\ee

This establishes that the partition function in a topological field
theory is independent of both the metric and coupling constants,
providing the theory remains well defined as we vary the parameters. Let
the parameter space be ${\cal T}$, which, for simplicity we take to be
connected. One can get from one set of
parameters $t_{i}$ to another $t_{i}'$ along some path in ${\cal T}$.
Pick such a path and suppose that for all points along
the path the theory is well behaved. Now perturb the path. If only for
very special choices of the perturbation does the path go through points
that lead to an ill defined theory, one says that the partition function
is independent of generic variations of the couplings. This is good
enough.

Unfortunately, it is not always the case that a generic path misses the
`bad' points, as we will see
later, there are situations in Donaldson and in Witten theory where the
dimension of the moduli space jumps as one varies the metric and that
one cannot avoid this.

%\vspace{.8cm}

\subsection{No Physical Degrees of Freedom}

When dealing with any field theory, there are certain restrictions on the
field content. For example, one should
have a good quadratic form. This means that, in the present situation, up
to a finite number of
zero modes, there should be an equal number of degrees of
freedom in the set of fields $\{ \Phi \}$ and $\{ B \}$, even though the
labels are different.
For example, the four dimensional Yang-Mills action (plus a $\theta$
term), may be written as
\be
\Tr \int_{\RR^{4}}\left( B^{\mu \nu}_{+}F_{\mu \nu} - \frac{g^{2}}{2}B^{\mu
\nu}_{+}
B_{\mu \nu}^{+} \right) ,
\ee
where $B_{\mu \nu}^{+}$ is a self dual anti-symmetric tensor. The equation
of motion for $B_{+}$ is algebraic, so that one can substitute this back
to obtain the more usual form of the action\footnote{In the Path
integral we are dealing with a Gaussian integral over $B$ which amounts
to the same thing. }. Now $B_{\mu \nu}$, only has three independent
components while $A_{\mu}$ has four. We need to gauge fix and we can do
so by introducing a multiplier field $b$ and Fadeev-Popov ghosts to the
theory. Now one adds
\be
\Tr \int_{\RR^{4}} \left( b \partial^{\mu}A_{\mu} + \ol{c}
\partial^{\mu} D_{\mu}c \right)
\ee
to the action. $B$ and $b$ together have four degrees of freedom
and so match the gauge field, while $c$ has one degree of freedom and
matches $\ol{c}$ so all is well.

This counting implies that we have a well defined action but does not
tell us what the {\em physical} degrees of freedom are. We know that in
$d$-dimensions a vector field has $d-2$ physical degrees of freedom.
There is a simple way of getting this from the gauge fixed action. A
gauge field has $d$ degrees of freedom, and each of the two ghosts has
$-1$ degrees of freedom, giving us a total of $d-2$. The $B$ and
$b$ fields do not count as they are `non-propagating', meaning one can
eliminate them algebraically.

The topological field theory has, by construction, for every field an
associated `ghost' field identical in every respect except that it has
opposite Grassmann parity. As every field is matched by a ghost, the
total number of physical degrees of freedom is always zero in a
topological field theory.

%\vspace{.8cm}

\subsection{Index theory and the Dimension of ${\cal M}$}

We have seen that there are no propagating degrees of freedom, we are
down to the zero, or topological, modes and some of these make up the
moduli space ${\cal M}_{s}$. What is the dimension of ${\cal M}_{s}$? To
simplify life let us assume that ${\cal M}_{s}$ is connected and smooth
about most points. One way to determine the dimension of ${\cal M}_{s}$,
is to fix a point $\f \in {\cal M}_{s}$ and then see in how many
directions you can go and stay in ${\cal M}_{s}$. We saw an example of
this argument in the lectures of J. Harvey for the moduli space of
monopoles and we proceed in the same way. Hence, we want $s(\f) = 0
$ and if $\f + \d \f$ is a nearby point we also require $s(\f + \d \f)
=0$, or we look for solutions to
\be
\left. \frac{\d s^{a}(\Phi)}{\d \Phi_{i}} \right|_{\f} . \d\f_{i} = 0 .
\label{defcomp}
\ee
If we are lucky the operator $D= \d s^{a}(\f)/\d \f$ has no co-kernel, as
in the case of the monopoles, and the index is known. Then as
${\textstyle index}(D) = {\textstyle Ker}(D) - {\textstyle CoKer}(D)$ we
would have the dimension of the moduli space. There are many situations
where we are not lucky. On a compact odd dimensional manifold the
operator $D$ that is associated with the space of flat connections
(\ref{flat}) has index zero (as the kernel and cokernel are equal). In
such situations one must look elsewhere for a handle on the moduli
space. For flat connections the equivalent description in terms of
homomorphisms of the fundamental group of the underlying manifold into
the gauge group (modulo the adjoint
action of the gauge group) contains the necessary information.

The index of $D$ is sometimes called the virtual dimension of the moduli
space.

\subsection{The Euler Character}\label{ecex}

One of the classic invariants of a closed compact n-manifold $X$ is its Euler
character. It is defined to be
\be
\chi(X)= \sum_{i=1}^{n}(-1)^{i}b_{i} \, \label{ec}
\ee
where the Betti numbers are $b_{i}= {\textstyle dim}H^{i}(X,\RR)$. Now
there are two well known formula for this invariant. The first, due to
Gauss and Bonnet states that if $R_{\mu \nu \kappa \lambda}$ is the
Riemann curvature tensor then (with $R$ the curvature two form)
\be
\chi(X) = \frac{1}{(2\pi)^{n/2}}\int_{X}R^{n/2} \, . \label{GB}
\ee
The second formula, due to Poincar\'{e} and Hopf, counts the number of
non degenerate critical
points $x_{P}$ of a function $f$ on $X$ with sign,
\be
\chi(X) = \sum_{P} {\textstyle sign} \det{(\partial_{\mu} \partial_{\nu}
 f)}\, . \label{PH}
\ee
These formulae arise naturally in the context of supersymmetric quantum
mechanics. The aim there is to give a path integral representation of
the index of certain differential operators. The index of the de-Rham
operator offers a third representation of the Euler character. As Witten
showed \cite{witmorse} there is a twisting of the de-Rham operator $d$ that
interpolates between the two formulae, $d \rightarrow e^{-tf}de^{tf}$.
In the supersymmetric quantum mechanics path integral one can take the
limit $t \rightarrow 0$ to arrive at  (\ref{GB}) or $t \rightarrow
\infty$ to arrive at (\ref{PH}). By supersymmetry invariance the path
integral is formally independent of $t$ and the equality of the two
formulae is thus established.

This situation prompted Matthai and Quillen
\cite{MatQuil} to develop a
completely classical formula that interpolates between (\ref{GB}) and
(\ref{PH}). We will give a physicists `derivation' of this shortly.
Before doing that I would like to explain the historical relationship of the
Matthai-Quillen formalism to topological field theory. The
supersymmetric quantum mechanics, alluded to above, was, perhaps, the
first example of a Witten type topological field theory. After Wittens
introduction of Donaldson theory it was shown, by Atiyah and Jeffery
\cite{atJef},
that one could re-interpret the construction as an infinite dimensional
version of the Matthai-Quillen formalism. These infinite dimensional
Matthai-Quillen theories devolve to  finite dimensional Matthai-Quillen
models of the type we will presently discuss. In the quantum mechanics
context, the supersymmetric theory considered by Witten can be viewed as
an infinite dimensional Matthai-Quillen construction, while the finite
dimensional formula that it gives rise to and that interpolates between
(\ref{GB}) and (\ref{PH}) is the `classical' Matthai-Quillen
formula\footnote{ This is an example of the healthy feedback from
mathematics to physics
and vice-versa that we witness in this field}.

An account of this
construction in topological field theory is to be found in
\cite{btcasson,blau,cmr,vafwit}. Physicists will see that this
construction is equivalent
to the existence of a Nicolai map \cite{brt,mansfield,bbrt}.
%Thus we are in a
%kind of `do-loop' each feeding the other.

Now back to business.
Let $X$ be a closed and compact manifold, and $f$ a map, $f:X \rightarrow
 \RR $, which has isolated critical points. The points at which
\be
df =0
\ee
define our moduli space ${\cal M}_{f}$. If the critical points are
isolated this means that the second derivative of $f$ is not zero at
those points. From our discussion of the
dimension of the moduli space we should be looking for solutions to the
variation of $df$,
\be
\frac{D^{2}f}{Dx^{\mu}x^{\nu}} \d x^{\mu} =0 .
\ee
If the eigenvalues of $D^{2}f/Dx^{\mu}Dx^{\nu}$ are not zero then the
only solution is $\d x^{\mu} =0$, that is the critical point is isolated.

The supersymmetry algebra is
\bea
& & Qx^{\mu}= \psi^{\mu} \, , \; \; \; Q\psi^{\mu} = 0 \, , \nonumber \\
& & Q\pb_{\mu} = B_{\mu} - \pb_{\nu}\Gamma^{\nu}_{\mu\kappa} \psi^{\kappa}\, ,
 \nonumber \\
& &
QB_{\mu} = B_{\nu}\Gamma^{\nu}_{\mu \kappa}-\frac{1}{2}\pb_{\nu}
R^{\nu}_{\mu\lambda\kappa} \psi^{\lambda}
\psi^{\kappa} \, .
\eea
with $Q^{2}=0$.
Now we may create a topological (field) theory with the
action
\bea
S_{f} &= &  \{ Q, \pb_{\mu} g^{\mu \nu}(it\partial_{\nu}f + \frac{1}{2}
B_{\nu}) \} \nonumber \\
  &=& i tB^{\mu}\partial_{\mu}f + \frac{1}{2} g_{\mu \nu} B^{\mu}B^{\nu} -
it\pb^{\mu}\frac{D^{2}f}{Dx^{\mu}
Dx^{\nu}}\psi^{\nu} -\frac{1}{4}R_{\mu \nu \kappa \lambda}\pb^{\mu}
\psi^{\kappa}  \pb^{\nu} \psi^{\lambda} \, . \label{action}
\eea
The transformation rules have been chosen to give us a covariant action.

Notice that there is a second supersymmetry that one gets by exchanging
$\psi \rightarrow \overline{\psi}$ and $\overline{\psi} \rightarrow
-\psi$. This happens quite naturally whenever one wishes to write down a
topological field theory for an Euler character. The correspondence
comes because on the space of forms $Q$ acts like $d$ while the second
supersymmetry charge $\overline{Q}$ behaves as $d^{*}$. The construction
is most easily understood in terms of supersymmetric quantum mechanics,
which, unfortunately, there is no time to go into here.

The partition function
\be
Z_{f} = -\frac{1}{(2\pi i)^{n}}\int_{X} \int dB  d\psi d\pb e^{-S_{f}}
\label{EC}
\ee
is independent of smooth deformations of the parameters, as we discussed
previously, since the derivative of the action with respect to either
the metric or $t$ is of the form $\{ Q, ...\}$ (we see this directly from
the first line of (\ref{action})). In particular, it does not depend on
$t$, so that we are free to take various limits.

\vspace{.8cm}
\noindent \underline{$t \rightarrow \infty$}

\noindent In this case the entire contribution to the `path integral' is
around the critical points of $f$. To see this send $B\rightarrow
\frac{1}{t}B$ and $\pb \rightarrow \frac{1}{t}\pb$,
\be
S_{f}(\infty) = ig^{\mu \nu}B_{\mu}\partial_{\nu}f  -i g^{\mu
\nu}\pb_{\mu} \frac{D^{2}f}{Dx^{\nu}
Dx^{\lambda}}\psi^{\lambda} \, .
\ee
An important feature of this scaling is that the Jacobian of the
transformation is unity.

We let $\{x_{P}\}={\cal M}_{f}$
be the critical point set and expand around each point as $x=x_{P}+
x_{q}$. Furthermore, around a critical point we can pick a flat metric
$g_{\mu \nu}= \d_{\mu \nu}$ for
which the christoffel symbol vanishes. The integral over $B$, gives
\bea
(2\pi)^{n}\d\left(\partial_{\mu}f(x)\right)& =& (2\pi)^{n}\sum_{P}\d\left(
x^{\nu}_{q}
 \partial_{\mu}\partial_{\nu}f(x_{P}) \right) \nonumber \\
&=& (2\pi)^{n}\sum_{P} \frac{1}{|\det{\partial_{\nu}\partial_{\mu}f}|}\d \left(
x_{q}^{\nu} \right)
\eea
On the other hand the integral over the fermionic fields,
\be
\int d\psi d\pb \exp{\left(-i \pb^{\nu}\frac{\partial^{2}f}{\partial x^{\nu}
\partial
x^{\mu}} \psi^{\mu}\right)} = -(i)^{n}\det{\partial_{\nu}\partial_{\mu}f} \, .
\ee

The path integral, therefore becomes
\be
\int_{X}\sum_{P}
\frac{\det{\partial_{\nu}\partial_{\mu}f}}{| \det{\partial_{\nu
}\partial_{\mu}f}|}\d \left(
x_{q}^{\nu} \right) = \sum_{P}
\frac{\det{\partial_{\nu}\partial_{\mu}f}}{|\det{\partial_{\nu
}\partial_{\mu}f}|} \, .
\ee

We can write this in the form
\be
Z_{f} = \sum_{P} \eps_{P} \label{ph}
\ee
with
\be
\eps_{P}= {\textstyle sign} \det{H_{P}(f)}
\ee
with $H_{P}f = \partial^{2}f/\partial x^{\mu} \partial x^{\nu}$. $H_{P}f$ is
called the Hessian of $f$ at $x_{P}$.

\vspace{.8cm}

\underline{Example: Riemann Surfaces}

Consider the example of a height function of a genus $g$ Riemann surface as
given in figure 1.
\begin{center}
\fbox{Figure 1}
\end{center}

The critical points of the height function are
marked. The bottom of the surface, $f= h_{min}$, is a minimum, and hence the
sign of both eigenvalues of $\partial^{2}f/\partial x^{\mu} \partial
x^{\nu} $ are plus. The Hessian is therefore $+1$. At the top of the
surface, $f=h_{max}$, is a maximum, both eigenvalues are negative, the
determinant is positive, and the Hessian is $+1$. For each hole (there
are $g$ of them) one has two turning points. The turning points are at
$f= h_{1}, \dots ,h_{2g}$. One of the eigenvalues is positive, the other
negative so, at each of these points, the Hessian is $-1$.

All in all we obtain in this way,
\be
Z=2-2g
\ee
which we recognise as the Euler character of the Riemann surface.

\vspace{.8cm}

\noindent \underline{$t \rightarrow 0$}

In this limit we are left only with the curvature term, so that
\be
Z= -\frac{1}{(2\pi i)^{n}}\int_{X} \int dB e^{-\frac{1}{2}g_{\mu \nu}
B^{\mu} B^{\nu} } \int d\psi
d\overline{\psi} e^{\frac{1}{4}R_{\mu \nu \kappa \lambda}
\overline{\psi}^{\mu}\psi^{\kappa}\overline{\psi}^{\nu}\psi^{\lambda}}
\ee
The integrals over the fields $B$, $\pb$ and $\psi$ are now easy to
perform, leaving us with the Gauss-Bonnet formula for the Euler character
of a manifold $X$,
\be
\chi(X) = \frac{1}{(2\pi)^{n/2}} \int_{X} R^{n/2} \label{gb}
\ee

\vspace{.8cm}

\noindent \underline{Perturbations}

The form of the partition function (\ref{ph}) appears to depend quite
strongly on the function $f$ that we started with. Yet, topological (BRST)
invariance allowed us to equate this with the form of the partition
function as given in (\ref{gb}), which does not depend at
all on the function $f$ that we started with. There is a nice way to see
why this might be so. Consider a perturbed height function $f'$ as
displayed in figure 2.
\begin{center}
\fbox{Figure 2}
\end{center}

The difference between this and $f$ of figure 1 is
the addition of a `hill' and of a valley. Now, the apex of the new hill
is a maximum, so the Hessian there is $+1$. On the other hand the bottom
of the valley is a turning point with Hessian $-1$. So we see that the
addition of the Hessians for $f'$ at these two critical points cancel out,
and the sum reverts to that of the other critical points where $f'$ agrees
with $f$! One can easily convince oneself that whenever a valley is added then
so is a hill (when you dig a hole
you get a mound of dirt) as well as the converse, and the contributions
of the Hessian always cancel out .

This takes care of the undulations but what happens when we hit a plateau?
Such a situation is depicted in figure 3 for a height function $g$.

\begin{center}
\fbox{Figure 3}
\end{center}

This
situation means that the moduli space, i.e. the solution set to $dg=0$, is
not made up of just isolated points. In the
current situation there are
two ways out. The first, is simply to note that the general formula
(\ref{EC}), works in this instance as well. Indeed there are two
different limits that can be used. One can take the $t \rightarrow
0$ limit without fear. Alternatively, away from the plateau, one may use the
the $t\rightarrow \infty$ limit, and as one approaches the plateau, revert
to the $t\rightarrow 0$ limit. The second way to proceed, which will be
of importance later, is to perturb the function $g$ to a new function $g'$.
The perturbation $g'-g$ need only be ever so slight and then the critical
points are isolated again. The perturbed equation is
\be
dg = \eps v \label{PA}
\ee
where $v$ is a vector field (essentially $dg'-dg$). As long as $\eps >0$
there are only isolated solutions to this equation. One can construct a new
action which takes its values at (\ref{PA}),
\be
\{ Q , \overline{\psi}^{\mu}(\partial_{\mu}g - \eps v_{\mu}) \} \, ,
\ee
(all reference to the metric has been dropped, as it plays no role here,
we have taken the $t\rightarrow \infty$ limit).
The partition function function once more, gives the Euler character of the
surface, and the $\eps \rightarrow 0^{+}$ may be taken with impunity.

\subsection{Invariants and Zero Dimensional Moduli Spaces}

Quite generally (up to certain compactness requirements) given a system
of equations
\be
s^{a}(\Phi)=0
\ee
with isolated solutions $\{ \f \}$, the signed sum of solutions
\be
\sum_{\f}\eps_{\f} \label{altsum}
\ee
where $\eps_{\f} = {\textstyle sign} \det{( \d^{2}s/\d\f\d\f)}$, is a
topological invariant.  We have given a path integral proof of this for
the Euler character.

The equations which will occupy us for most of the
following lectures have isolated solutions. We do not really need all
the machinery of topological field theory once we have established that
the moduli space is zero dimensional. The sum (\ref{altsum}) will
be guaranteed to be a topological invariant. For this reason I will not
bother to write down a topological action corresponding to the monopole
equations (see section 6). However, as there are some extra complications
in writing
down a topological gauge theory action I will generalize the proceeding
discussion to that case. The examples that I will give are of Abelian and
non-Abelian instantons.

The catch in the case of gauge theories, when
the dimension of the moduli space is zero or if it has many components, is in
determining the signs in (\ref{altsum}). In the case of the solutions
to the monopole equations Witten has given a prescription for
fixing signs.

%\begin{center}
%\fbox{To be continued}
%\end{center}

\section{A Mathematical Digression}
The rest of these notes are concerned with four-dimensional field
theories living on a four manifold $X$.

\subsection{Homology and Cohomology Groups}
In this section I review, very
briefly and rather heuristically, some of
the constructs that we will be using repeatedly later on. As some of the
basic invariants of a smooth manifold are the homology groups (and
equivalently over $\RR$ the cohomology groups) we start there. A fine review
of this material, aimed at physicists, is \cite{egh}. From now on
$X$ is a smooth, compact, oriented and connected manifold.

\vskip .8cm
\noindent\underline{ {\bf Homology}}

Given a set of $p-$dimensional oriented submanifolds $N_{i}$ of $X$ one
can form a $p-$chain.

\noindent{\bf Definition:} A $p-$chain $a_{p}$ is
\be
\sum_{i}c_{i}N_{i} \, ,
\ee
where the coefficients $c_{i}$ determine the type of chain one has. If
the $c_{i} \in \RR (\CC)$ the $p-$chain is said to be a real (complex)
$p-$chain. For $c_{i} \in \ZZ (\ZZ_{2})$, and the chain is called an
integer ($\ZZ_{2}$) chain and so on.

$\pd$ is the operation that gives the oriented
boundary of the manifold it acts on. For example, the sphere has no
boundary so $\pd S^{2} = \emptyset $,
while the cylinder, $I \times S^{1}$, has two circles for a boundary,
$\pd (I \times S^{1})=\{0\} \times S^{1}\oplus \{1\} \times -S^{1} $. I
have written $\oplus$ to indicate that we will be `adding' submanifolds
to form chains when one should have used the union symbol $\cup$. The
minus sign in front of the second $S^{1}$ is to indicate that it has
opposite relative orientation. One defines the boundary of a $p-$chain
to be a $(p-1)-$chain by
\be
\pd a_{p} = \sum_{i}c_{i}\pd N_{i} .
\ee
Notice that $\pd^{2}= 0$, as the boundary of a boundary is empty.

\noindent{\bf Definition:} A $p-$cycle is a $p-$chain without boundary,
i.e. if $\pd a_{p}=0$ then $a_{p}$ is a $p-$cycle.

\noindent{\bf Definition:} Let $Z_{p}= \{ a_{p}: \pd a_{p}=0 \}$ be the
set of $p-$cycles and let $B_{p}= \{ \pd a_{p+1} \}$ be the set of
$p-$boundaries of $(p+1)-$chains. The $p-$th simplicial homology group
of $X$ is defined by
\be
H_{p} = Z_{p}/B_{p} .
\ee

\vskip .8cm
\noindent\underline{ {\bf Cohomology} }

\noindent{\bf Definition:}Let $Z^{p}$ be $\{ \o_{p}:d\o_{p}=0 \}$ the
set of closed $p-$forms and let $B^{p}$ be $\{\o_{p}: \o_{p} = d\o_{p-1}
\}$ the set of exact $p-$forms. The $p-$th De Rham cohomology group is
defined by
\be
H^{p}= Z^{p}/B^{p} .
\ee

\vskip .8cm
\noindent\underline{ {\bf De Rham's Theorems}}

\noindent{\bf Definition:} The inner product of a $p-$cycle, $a_{p}\in
H_{p}$ and a closed $p-$form, $\o_{p}\in H^{p}$ is
\be
\pi(a_{p},\o_{p}) = \int_{a_{p}}\o_{p}.
\ee
Notice that this does not depend on the representatives used for, by Stokes
theorem,
\bea
\pi(a_{p} + \pd a_{p+1}, \o_{p})&  = &\int_{a_{p}+ \pd a_{p+1}} \o_{p} =
\int_{a_{p}}\o_{p} + \int_{a_{p+1}}d\o_{p} = \pi(a_{p},\o_{p}) \nonumber
\\
\pi(a_{p},\o_{p}+d\o_{p-1}) &=& = \int_{a_{p}}\o_{p}+ \int_{\pd
a_{p}}\o_{p-1} = \pi(a_{p},\o_{p}) .
\eea

One may thus think of the {\bf period} $\pi$ as a mapping
\be
\pi:H_{p}(X,\RR)\otimes H^{p}(X,\RR) \rightarrow \RR
\ee
For $X$ compact and closed De Rham has established two important theorems.
Let $\{ a_{i}\}$, $i=1, \dots b_{p}$, be a set of independent
$p-$cycles forming a basis $H_{p}(X,\RR)$, where the $p-$th Betti number
$b_{p}= $dim$H_{p}(X,\RR)$.

\noindent{\bf Theorem 1:} Given any set of periods $\nu_{i}$, $i=1 \dots
b_{p}$, there exists a closed $p-$form $\o$ for which
\be
\nu_{i}= \pi(a_{i},\o)= \int_{a_{i}}\o .
\ee

\noindent{\bf Theorem 2:} If all the periods of a $p-$form $\o$ vanish,
\be
\pi(a_{i},\o)=\int_{a_{i}} \o =0
\ee
then $\o$ is an exact form.

Putting the two theorems together, we have that if $\{  \o_{i}\}$ is a
basis for $H^{p}(X,\RR)$ then the period matrix
\be
\pi_{ij}=\pi(a_{i},\o_{j})
\ee
is invertible. This implies that $H^{p}(X,\RR)$ and $H_{p}(X,\RR)$ are
dual to each other with respect to the inner product $\pi$ and so are
naturally isomorphic.

\vskip .8cm
\noindent\underline{{\bf Poincar\'{e} Duality}}

If $X$ is compact, orientable and closed of dimension $n$ then
$H^{n}(X,\RR) = \RR$, as, up to a total differential, any $\o_{n}\in
H^{n}(X,\RR )$ is proportional to the volume form. One aspect of
Poincar\'{e} duality is the

\noindent{\bf Theorem} $H^{p}(X,\RR)$ is dual to
$H^{n-p}(X,\RR)$ with respect to the inner product
\be
(\o_{p},\o_{n-p}) = \int_{X}\o_{p}\o_{n-p}.
\ee
This implies that $H^{p}(X,\RR)$ and $H^{n-p}(X,\RR)$ are isomorphic as
vector spaces, so that, in particular, $b_{p}=b_{n-p}$.

We need the following statement

\noindent{\bf Theorem:} Given any $p-$cycle $a_{p}$ there exists an
$(n-p)-$form $\a$, called the Poincar\'{e} dual of $a_{p}$, such that for
all closed $p-$forms $\o$
\be
\int_{a_{p}}\o = \int_{X}\a \o .
\ee

\vskip .8cm
\noindent\underline{{\bf The K\"{u}nneth Formula}}

This is a formula that relates the cohomology groups of a product space
$X_{1} \times X_{2}$ to the cohomology groups of each factor. The
formula is
\be
H^{p}(X_{1}\times X_{2},\RR) = \sum_{q=0}^{p}H^{q}(X_{1},\RR) \otimes
H^{p-q}(X_{2}\RR) .
\ee
In particular this implies that $b_{p}(X_{1}\times X_{2})=
\sum_{q+r=p}b_{q}(X_{1}) b_{r}(X_{2})$.

\vskip .8cm
\noindent{\bf Notation:} I will denote a basis for $H_{1}(X,\RR)$ by
$\ga_{i}$ and the dual basis for $H^{1}(X,\RR)$ by $[\ga_{i}]$. For
$H_{2}(X,\RR )$ I denote the basis by $\S_{i}$ (as a mnemonic for a
two dimensional manifold) and the corresponding basis for $H^{2}(X,\RR)$
by $[\S_{i}]$.

\subsection{Hodge Theory}
The De Rham theorems are very powerful and very general. Rather than
working with a cohomology class $[\o_{p}]$ of a $p-$form $\o_{p}$ it would be
nice to be able to choose a canonical representative of the class. This
is what Hodge theory gives us and along the way also gives us a different
characterisation of the cohomology groups.

Up till now, we have not needed a metric on the manifold $X$. The
introduction of a metric, though not needed in the general framework,
leads to something new. Let $X$ be equipped with a metric $g$. Define
the Hodge star $*$ operator, $* : \Omega^{p} \rightarrow
\Omega^{(n-p)}$, by
\be
* \o_{\mu_{1}, \dots , \mu_{p}} dx^{\mu_{1}}\dots dx^{\mu_{p}} =
\frac{\sqrt{g}}{(n-p)!} \o_{\mu_{1}, \dots , \mu_{p}} \eps^{\mu_{1} \dots
\mu_{p}}_{\; \; \; \; \; \; \mu_{p+1} \dots \mu_{n}} dx^{\mu_{p+1}}\dots
dx^{\mu_{n}} .
\ee
The symbol $\eps_{\mu_{1} \dots \mu_{n}}$ is $0$ if two labels are
repeated and $\pm$ for even or odd permutations respectively. All labels
are raised and lowered with respect to the metric.

\noindent\underline{ {\bf Theorem:}} For a compact manifold without
boundary, any $p$-form can be uniquely decomposed as a sum of an exact,
a co-exact and a harmonic form,
\be
\o_{p} = d\a_{(p-1)} + \d \b_{(p+1)} + \ga_{p}
\ee
this is referred to as the Hodge decomposition.

The harmonic form $\ga_{p}$ is our representative for $[\o_{p}]$. This
corresponds to choosing $d*\o_{p}=0$ as the representative. From the
point of view of gauge theory, this amounts to the usual Landau gauge
(extended to higher dimensional forms).

\noindent\underline{ {\bf Self-Dual Forms on a Four-Manifold}}

The Hodge star operator $*$ squares to unity,
$*^{2}=1$, when it acts on even forms. A consequence of this is that one
can, in four dimensions, orthogonally split the
space of two-forms according to
\bea
\o_{2} &=& \frac{1}{2}(1+*)\o + \frac{1}{2}(1-*)\o \nonumber \\
  &=& \o_{+} + \o_{-}
\eea
as $(1 \pm *)/2$ are projection operators. Still in four dimensions, if a
two-form $\o \in \Omega^{2}(X,\RR)$ is self dual
then we may refine the Hodge decomposition somewhat. Let
\be
\o = dA + * dB + [\S ]
\ee
where $A$ and $B$ are one-forms and $[\S ]$ is a harmonic form. Anti-self
duality, or $ (1+*)\o =0$ means that
\be
\o = -*\o = -dB -*dA - *[\S]
\ee
or, as the Hodge decomposition is unique, that
\be
A= -B \, , \; \; \; *[\S] = -[\S] .
\ee
This means that we can write for any (anti) self dual two-form
\be
\o = (1-*) dA + [\S]
\ee
with $* [\S] = -[\S]$.

We may also split the cohomology group $H^{2}(X)$, as a vector space, into
the two orthogonal pieces
\be
H^{2}(X)=H^{2}_{+}(X) \oplus H^{2}_{-}(X) .
\ee
Notice that the definition of the splitting depends on the metric that
is used to define $*$. One would get a different splitting if one used
another metric $*'$, however, the dimensions of the spaces $H^{2}_{+}(X)$
and $H^{2}_{-}(X)$ do not depend on the choice of metric. Denote the
dimensions by $b_{2}^{+}$ and $b_{2}^{-}$ respectively.

\subsection{Intersection Form}

Just as in two dimensions, there are surfaces in our 4-manifold that
cannot be contracted to a point. A simple example is that of $X= S^{2}
\times S^{2}$, it is clear that neither of the $S^{2}$ factors is
contractable to a point. The homology groups $H_{p}(X,\RR)$ are a
measure of noncontractible surfaces of dimension $p$ in $X$.
Let $X$ be a compact, oriented simply connected 4-manifold. If $\a$
represents a class in $H_{2}(X,\ZZ)$ then, by Poincar\'{e} duality, we can
identify it with a class in $H^{2}(X,\ZZ)$, which I will also denote by
$\a$. Given $\a,\b \in H_{2}(X,\ZZ)$ we can define a quadratic form, (a
$b_{2} \times b_{2}$ matrix),
\be
Q: H_{2}(X;\ZZ) \times H_{2}(X,\ZZ) \rightarrow \ZZ ,
\ee
by
\be
\int_{X} \a \b \equiv \a.\b \, .
\ee

$Q$ enjoys the following properties.
\begin{itemize}
\item It is unimodular ($\det{Q}=\pm 1$). This follows from the fact that
it provides
the Poincar\'{e} duality isomorphism between $H_{2}(X)$ and $H^{2}(X)$.
\item It is symmetric. This is a trivial consequence of the fact that
two forms commute, i.e. $\a \b = \b \a$, whenever either of $\a$ or $\b$
$\in H^{2}(X,\RR)$.
\end{itemize}

\noindent\underline{\bf Examples:}
\begin{enumerate}
\item The four sphere $S^{4}$ has trivial $H_{2}$ so all the
intersection numbers vanish.
\item Let $X=S^{2} \times S^{2}$. There are two basic two forms,
$\o_{1}$ and $\o_{2}$ dual to the second and first $S^{2}$ respectively.
We have
\be
Q(\o_{1},\o_{2})=\int_{S^{2}\times S^{2}}\o_{1}\o_{2}= \int_{S^{2}\times
\{ p\}} \o_{1} = \int_{
\{ p\} \times S^{2}} \o_{2} = 1 .
\ee
With this we see that
\be
Q= \left( \begin{array}{cc}
0 & 1 \\
1 & 0
\end{array}\right) .
\ee
\item Consider a product $4$-manifold, $X = \S_{1}\times
\S_{2}$, where the $\S_{i}$ are Riemann surfaces of genus
$g_{i}$. In this case
\be
Q = \sum_{2g_{1}g_{2}+1} \oplus \left(\begin{array}{cc}
0 & 1 \\
1 & 0
\end{array}\right) .
\ee
\item The manifold $\CC \PP^{2}$, has $b_{2}=1$, so $Q=1$.
\item The $K_{3}$ surface has $b_{2}= 22$! This means that $Q$ is a
$22 \times 22$ matrix. Indeed,
\be
Q = 3\left( \begin{array}{cc}
0 & 1 \\
1 & 0
\end{array} \right) \oplus 2(-E_{8}) ,
\ee
where $E_{8}$ is the Cartan matrix of the exceptional Lie algebra
$e_{8}$,
\be
E_{8} = \left( \begin{array}{cccccccc}
2 &-1& 0  & 0 & 0&0 &0 &0 \\
-1& 2 &-1  & 0 &0 &0 &0 &0  \\
0 & -1 & 2 & -1 & 0 & 0 & 0 & 0 \\
0 & 0& -1 & 2 & -1 & 0 & 0&0 \\
0 & 0 & 0 & -1 & 2 & -1 & 0 & -1 \\
0 & 0 & 0 & 0 & -1 & 2 & -1 & 0 \\
0 & 0 & 0 & 0 & 0 & -1 & 2 & 0 \\
0 & 0 & 0 & 0 & -1 & 0 & 0 & 2
\end{array} \right)
\ee

\end{enumerate}
%\begin{center}
%\fbox{To be continued}
%\end{center}

The intersection form $Q$ is the basic invariant of a compact
four-manifold. It will feature prominently in the next sections.

\section{S Duality In Maxwell Theory and Abelian Instantons}
In this section we will study two different aspects of four-dimensional
Abelian theories. The first is a study of $S$ duality in the Abelian
context.  M.
Bershadsky has described, in his lectures, a relationship between $S$
duality in four dimensions and $T$ duality in two dimensions, for
non-Abelian theories. Why the interest in the Abelian case? From the
lectures of Harvey and
Alvarez-Gaume we have seen the important part role played by the
breaking of $SU(2)$ down to $U(1)$ in the $N=2$ supersymmetric
Yang-Mills theory. The effective theory, at strong coupling is a $U(1)$
theory.

There have appeared in the last month two very interesting papers on $S$
duality in Maxwell theory, which go in different directions. E. Verlinde
\cite{versdual} has studied the $S$ duality of Maxwell theory and its
relationship to $T$ duality, string theory and higher dimensional (free)
field theories. E. Witten \cite{witsdual} has used it to probe the
modular properties of the partition function so as to fix some of the
$\tau$ dependence of the $N=2$ theory on arbitrary four manifolds.

In the following I will give a `bare bones' description of $S$ duality
for Maxwell theory which I hope, though it does no justice to the above
works, will nevertheless entice the reader to look into the references.

The second subject will be a quick tour of Abelian instantons and the
construction of a topological field theory that describes the moduli
space. That model should not be taken too seriously- I have included it
so as to explain how a topological field theory may contain `$\tau$'
dependence and to introduce some ideas that we will need later on.

\subsection{Maxwell Theory on $X$ and $S$ Duality}
The usual action for pure Maxwell theory with a theta term is
\be
S = \frac{1}{g^{2}}\int_{X} F_{A} *F_{A} +i \frac{\theta}{8\pi^{2}} \int_{X}
F_{A}F_{A} . \label{maxact}
\ee
There is always a little confusion with the field strength. A gauge field
$A$ is denoted by
\be
A= A_{\mu}dx^{\mu}
\ee
and it is natural to define the field strength as $F=dA$. However, in
components we find
\bea
F = dA &=& \partial_{\mu}A_{\nu}dx^{\mu}dx^{\nu} \nonumber \\
       &=& \frac{1}{2}\left( \partial_{\mu}A_{\nu}-\partial_{\nu}
    A_{\mu} \right) dx^{\mu}dx^{\nu} \nonumber \\
  & = & \frac{1}{2}F_{\mu\nu}dx^{\mu}dx^{\nu}
\eea
so that there is a factor of $1/2$ in the definition of the components.
With these conventions under control we deduce that
\be
\int_{X}F*F = \frac{1}{2}\int_{X} d^{4}x \sqrt{g} F_{\mu \nu}F^{\mu \nu}
\, , \; \; \;
\int_{X}FF = \frac{1}{4}\int_{X} d^{4}x \sqrt{g} \eps_{\a \b \mu \nu}
F^{\a \b}F^{\mu \nu} .
\ee

The partition function, with action (\ref{maxact}), is a function of both
$\tau$ and
$\overline{\tau}$ where
\be
\tau = \frac{\theta}{2\pi} + i \frac{4\pi}{g^{2}} .
\ee
What we would like to know is how the partition function behaves under the
action of $SL(2,\ZZ)$ on $\tau$. That action is described by
\be
\tau \rightarrow \frac{a\tau+ b}{c\tau +d} \label{sl2z}
\ee
where the constants $a,b,c,d$ are integers and obey $ad-bc=1$ (so that
one may group them together into a matrix
\be
 \left( \begin{array}{cc}
a & b \\
c & d
\end{array} \right)
\ee
which clearly defines the group $SL(2,\ZZ)$ ). One can generate
$SL(2,\ZZ)$ by the transformations
\be
S= \left( \begin{array}{cc}
0 & 1 \\
-1 & 0
\end{array} \right)
\ee
and
\be
T=\left( \begin{array}{cc}
1 & 1 \\
0 & 1
\end{array} \right)
\ee
{}From these we see that $T(\tau)= \tau +1$ (just substitute into
(\ref{sl2z}) the values $a=b=d=1$ and $c=0$) or $\theta \rightarrow
\theta + 2\pi$. If the partition function is invariant under $T$ then we
are saying the physics is periodic in $\theta$ with period $2\pi$.
Likewise the action of $S$ on $\tau$ is $S(\tau)= -1/\tau$, or
$g^{2}/4\pi \rightarrow 4\pi/g^{2}$. The label $S$ is thus apt for it
has the effect of exchanging weak and strong coupling.

It is strange to talk about weak and strong coupling for a free theory!
The context in which strong-weak duality has been discussed in the school
is in theories with monopoles. Now one can mimic the presence of
monopoles by allowing for manifolds with non trivial two-cycles. The
`Dirac quantization' condition applies or, put another way, the integral
of $F_{A}$ over such a surface must be $ 2\pi$ times an integer.
A small surprise is that for some four manifolds the
partition function transforms well under $T$ and $S^{2}$ but not under
$S$. We will find that the partition function is a modular form of particular
weight (see below), of $SL(2,\ZZ)$ or a (finite index) subgroup thereof.

We can now check for the properties of the partition function under both
$\tau \rightarrow \tau +\a$ and $\tau \rightarrow -1/\tau$.

\vspace{.8cm}

\noindent\underline{$\tau \rightarrow \tau + 1$ or $\tau \rightarrow
\tau + 2$}

$\tau \rightarrow \tau + \a$ is
easy to check as it corresponds to $\theta \rightarrow \theta + 2\pi \a$.
The action shifts by
\be
i\frac{\a}{4\pi}\int_{X}F_{A}F_{A} . \label{c2}
\ee
Now we know that $F_{A} = da + 2\pi n^{i}[\S_{i}]$ where $a \in
\Omega^{1}(X,\RR)$ and $[\S_{i}]$ is a basis of $H^{2}(X,\RR)$ dual to a
basis of $H_{2}(X,\RR)$, i.e. $\int_{\S_{j}}[\S_{i}]= \d_{ij}$, and that
the $n^{i}$ are integers. This
decomposition is due to the, by now, familiar magnetic flux condition
\be
\int_{\S_{i}}F_{A}= 2\pi n^{i} .
\ee
With these conventions (\ref{c2}) becomes
\be
i\pi \a n^{i}Q_{ij}n^{j}
\ee
where $Q_{ij}= \int_{X} [ \Sigma_{i}] [\Sigma_{j}]$. More on the matrix
$Q$ later.
If $n.Q.n$ is even then $\exp{(-S)}$ is unchanged for $\a =1$, $\theta
\rightarrow \theta + 2\pi$, while if
$n.Q.n$ is odd one needs to take $\a=2$, $\theta
\rightarrow \theta + 4\pi$, to have an invariance. There are general
results that tell us that for manifolds on which fermions are defined
$n^{2}$ must be even. One says that there is a spin structure. For
manifolds which do not admit a spin structure there is no condition on
$n^{2}$. Lets turn to some examples.

\begin{itemize}

\item $X=\Sigma_{1}\times\Sigma_{2}$

We know that we can have spinors on Riemann surfaces so in this case we
would expect that $n.Q.n$ is even. First we have to decide what the
available harmonic two-forms are. We can use the Kunneth formula
$H^{p}(X_{1}\times X_{2}) = \sum_{q=0}^{q=p}\oplus H^{(q)}(X_{1}) \otimes
H^{(p-q)}(X_{2})$ and set $p=2$. To simplify matters I will do the case of $X=
S^{2}\times S^{2}$ and leave the general case as an exercise. Let us
denote the basic two forms by $\o_{1}$ and $\o_{2}$ of the first and the
second $S^{2}$ respectively. Kunneth tells us that these two span the
second cohomology group of $S^{2}\times S^{2}$. The $\o_{i}$ are
normalized by
\be
\int_{S^{2}_{i}}\o_{j}= \d_{ij}
\ee
and the only non-zero components of $Q$ are
\be
Q_{12}=Q_{21}= \int_{S^{2}\times S^{2}} \o_{1} \o_{2} = 1 .
\ee
Then we have
\be
n.Q.n = 2n^{1}n^{2}
\ee
which is even, as promised.

\item $X=\CC\PP^{2}$

The second cohomology group of $\CC\PP^{2}$ has only one generator which
we denote by $\o$. The harmonic part of $F_{A}$ is therefore
proportional to $\o$
\be
F_{A} = 2\pi n \o
\ee
and consequently $n.Q.n = n^{2}$ which can be even or odd. When
discussing Witten theory we will see that indeed $\CC\PP^{2}$ does not
admit a spin structure.
\end{itemize}

\vspace{.8cm}

\noindent\underline{$\tau \rightarrow -1/\tau$}

In order to check the modular properties of the partition function under
$\tau \rightarrow -1/\tau$ we need to re-write the action. Consider
instead of (\ref{maxact}) the action
\be
S(F,V)=\frac{1}{4\pi}\int_{X}F(i{\textstyle Re}\tau + {\textstyle Im}\tau *)F
-\frac{i}{2\pi}\int_{X}dV F .
\ee
Here $V$ is to be understood as a connection on a non-trivial bundle,
in defining the path integral we should sum over all such lines, and $F$
is an arbitrary two-form. We can
write
\be
F_{V}=dV = dv + 2\pi m^{i}[\Sigma_{i}] .
\ee

Firstly lets establish that the theory defined in this way is equivalent
to Maxwell theory, also summed over all non-trivial line bundles. The
partition function is
\be
Z(\tau,\overline{\tau}) = \sum_{m^{i}}\int DF Dv\, e^{-S(F,V)} .
\ee
$F$ may be decomposed as $F= *dB + da + c^{i}[\Sigma_{i}]$ where the
$b^{i}$ are real numbers. The integral over $v$ gives a delta function
constraint setting $dF=0$, which implies $B=0$. We are left with the sum
\be
\sum_{m^{i}} \exp{\left( i m^{i} Q_{ij} c^{j} \right) }
\ee
which is a periodic delta function which forces $c^{j}= 2\pi n^{j}$.
So the requirement that $F$, after integrating out $V$, is $F_{A}$ fixes
the coefficient of the $dVF$ term in the action (up to a sign). This
establishes that the partition function agrees with the Maxwell
partition function.

Now we integrate out $F$ instead of $V$. This is a simple Gaussian
integral (and I leave it as an exercise) giving
\be
Z(\tau,\overline{\tau}) = \sum_{m^{i}}\int Dv \exp{ \left(
\frac{-1}{g^{'2}} \int_{X}F_{V}*F_{V} -i\frac{\theta'}{8\pi^{2}}
\int_{X} F_{V}F_{V}\right) }
\ee
where
\be
\frac{4\pi}{g^{'2}} = \frac{{\textstyle Im}\tau}{\tau \overline{\tau}} ,
\; \; \; \frac{\theta'}{2\pi} = - \frac{{\textstyle Re}\tau}{\tau
\overline{\tau }} .
\ee
These equalities correspond to $\tau \rightarrow -1/\tau$. Notice that
we have not shown invariance of the partition function but rather
described its `covariance'.

In checking the duality transformations we have not at all worried about the
normalization of the path integral measure. We will certainly have to
worry about this if we want to be sure of the complete dependence on $\tau$ of
the partition function.

%\vspace{.8cm}

\subsection{`Semi-Classical Expansion'}

In order to perform the path integral over gauge fields with non-trivial
first Chern class (monopole number) we write the field strength as
\be
F_{A} = 2\pi n^{i}[\Sigma_{i}] + da \label{split}
\ee
and let the path integral be an integral over the globally defined
vectors $a$, though they may have non-trivial harmonic 1-form pieces, as
well as a summation over the $n^{i}$. It is worth remarking that the
split (\ref{split}) is a standard one we often employ in field theory,
namely a split into a classical configuration plus a quantum part. A
classical configuration in Maxwell theory is one that satisfies
\be
d*F_{A}=0
\ee
while all $F_{A}$ satisfy $dF_{A}=0$. This tells us that $F_{A} \in
H^{2}(X,\RR)$. The monopole quantization condition tells us that indeed
$F_{A}/2\pi \in H^{2}(X,\ZZ)$, so that classical configurations take the
form $F_{A} = 2\pi n^{i}[\Sigma_{i}]$.

Substituting (\ref{split}) into the action (\ref{maxact}) we see that
the classical and quantum configurations do not talk
\be
S=\frac{1}{g^{2}}\int_{X}da*da + \frac{4\pi^{2}}{g^{2}}n^{i}G_{ij}n^{j}
+i \frac{\theta}{2}n^{i}Q_{ij}n^{j}
\ee
where
\be
G_{ij} = \int_{X}[\Sigma_{i}]* [\Sigma_{j}]
\ee
is the metric on the space of harmonic two-forms. Consequently we may
split the path integral into a product of the $\tau$ dependent classical
part $Z_{c}$ and the Im$\tau$ dependent quantum part $Z_{q}$,
\be
Z(\tau)= Z_{c}(\tau) Z_{q}(\tau)
\ee
with
\be
Z_{c}(\tau) = \sum_{n^{i}} \exp{-\left( \frac{4\pi^{2}
}{g^{2}}n^{i}G_{ij}n^{j}
+i \frac{\theta}{2}n^{i}Q_{ij}n^{j} \right) } ,
\ee
and
\be
Z_{q}(\tau) = \int Da \, \exp{ \left( -\frac{1}{g^{2}}\int_{X}da*da
\right) } .
\ee
We see that, with some definition, the $\tau$ dependence of $Z_{q}$ is
only in the form of Im$\tau$. Witten gives us a way to fix this
dependence. The dependence is of the form
\be
Z_{q}(\tau) \sim \left(\sqrt{ {\rm Im}\tau }\right)^{b_{1}-1} .
\ee

Lets have a look at some simple examples.

\begin{itemize}
\item $X = \CC\PP^{2}$

We know that the classical part of $F_{A}= 2\pi n \o$. Furthermore, the
metric is such that $*\o =\o$. Hence
\be
Z_{c}(\tau)_{\CC\PP^{2}}= \sum_{n} \exp{ \left( -i \pi \overline{\tau} n^{2}
\right) } .
\ee
Notice that this is only invariant under $\tau \rightarrow \tau + 2$.
To discover the behaviour of $Z_{c}(\tau)_{\CC\PP^{2}}$ under $\tau \rightarrow
1/\tau$ we use a small trick. We rewrite $Z_{c}(\tau)_{\CC\PP^{2}}$ as
\be
\sum_{n \in \ZZ} \exp{ \left( -i \pi \overline{\tau} n^{2}
\right) } = \sum_{m \in \ZZ}\int_{-\infty}^{\infty}dy \exp{\left( -i
\pi y^{2} \overline{\tau} + 2\pi i y m \right) }
\ee
and the equality holds because the sum over $m$ will give zero unless $y$
is an integer. Now we can integrate out $y$ to obtain
\be
Z_{c}(\tau)_{\CC\PP^{2}} = \frac{1}{ \sqrt{i \overline{\tau}} } Z_{c}(
-\frac{1}{\tau})_{\CC\PP^{2}}
\ee
which tells us that $Z_{c}(\tau)_{\CC\PP^{2}}$ is a modular form of degree
$(0,1/2)$.

\item $X = \overline{\CC\PP}^{2}$

The only difference between $\CC\PP^{2}$ and $\overline{\CC\PP}^{2}$ is
an reversal of orientation, so that in this case $*\o = - \o$, and
\be
Z_{c}(\tau)_{\overline{\CC\PP}^{2}}= \sum_{n} \exp{ \left( -i \pi
\tau n^{2}
\right) } .
\ee
This is a modular form of degree $(1/2,0)$.

\item $ X =S^{2} \times S^{2}$

{}From our previous discussion we know that $F_{A} = 2\pi n^{1}\o_{1}
+ 2\pi n^{2}\o_{2} $. The only thing we need to specify is the metric $*$.
We do this by setting
\be
*\o_{1} = R^{2} \o_{2} \, , \; \; \; \; \; *\o_{2} = \frac{1}{R^{2}}
\o_{1} \, ,
\ee
which satisfies $*^{2}=1$. $Z_{c}$ for such configurations is
\be
Z_{c}(\tau)_{S^{2} \times S^{2}} = \sum_{n_{1} , n_{2} } \exp{ \left(
-i \frac{\pi}{2} \tau ( n_{1}R - \frac{n_{2}}{R})^{2} + i \frac{\pi}{2}
\overline{\tau} (n_{1}R + \frac{n_{2}}{R} )^{2}\right) }
\ee
which has a form familiar from the study of the $R \rightarrow 1/R$ symmetry
in string theory for a boson compactified on a circle of radius $R$
(what the partition function lacks are the propagating modes).

\end{itemize}

At this point there are various lines of investigation available. One is
to establish the $\tau$ dependence of $Z_{q}$; see \cite{witsdual}. On
the other hand the
similarity between the partition functions obtained here and those of
string theory lead one in another direction \cite{versdual}. Instead we
turn to a topological field theory.

\subsection{Abelian Instantons}
One may wonder what the space of solutions to the Abelian instanton
equation is. We would like to solve
\be
F_{\mu \nu}^{+} =0 \, . \label{abinst}
\ee
We know that, by the Bianchi identity, $dF=0$, and as
$F=*F$ this also implies $d^{*}F=0$, or that $F \in H^{2}(X, \RR)$.
Actually, as the flux of the gauge field through any two-surface is
quantized, $(F/2\pi) \in H^{2}(X,\ZZ)$. If $F$ is a solution to (\ref{abinst})
then $(F/2\pi) \in H_{-}^{2}(X, \RR)$.

One important observation is that the space of solutions {\em is} metric
dependent. Indeed on any $4$-manifold $X$ with $b_{2}^{+}>0$, after a
small perturbation of the metric, there are no Abelian instantons except
flat ones (so that if $X$ is simply connected as well the only solution
to (\ref{abinst}) is the trivial gauge field $A_{\mu}=0$). The reason
for this is that, as we have seen, $F$ must lie on a lattice, however it
also lives in $H_{-}^{2}(X,\RR)$, which will generically lie off the lattice.

To see how this could be so consider $S^{2}\times S^{2}$ with the two
generators of $H^{2}(X,\ZZ)$ described previously,
$b_{2}^{+}=b_{2}^{-}=1$. The metric was taken to satisfy
\be
*\o_{1} = \o_{2} \, , \; \; \; \; \; *\o_{2}=\o_{1} ,
\ee
so that $*^{2}=1$. In this case any $\o \in H^{2}(S^{2}\times
S^{2},\ZZ) $ can be expressed as
\bea
\o & =& n_{1}\o_{1} + n_{2}\o_{2} \nonumber \\
    & = & \frac{n_{1}+n_{2}}{2} \o_{+} +
\frac{n_{1}-n_{2}}{2}\o_{-} ,
\eea
where $\o_{\pm}=(\o_{1}\pm \o_{2})$ are the generators of
$H^{2}_{\pm}(S^{2}\times S^{2},\RR)$. From these expressions one sees
that $H_{-}^{2}$ intersects the lattice at $n_{1}=-n_{2}$. This is
displayed by the solid lines in figure 4.
\begin{center}
\fbox{Figure 4}
\end{center}

Now we are going to perturb the metric a little. Let
\be
*\o_{1} = \lambda \o_{2} \, , \; \; \; \; *\o_{2} =\frac{1}{\lambda}
\o_{1} ,
\ee
where $\lambda$ can be close to unity. With this metric the self-dual
anti-self dual basis is
\be
\o_{\pm} = \o_{1} \pm \lambda \o_{2} ,
\ee
and evidently $H^{2}(S^{2}\times S^{2},\ZZ) \cap H_{-}^{2}(S^{2}\times
S^{2},\RR) = (0,0)$. The
situation is summarised in figure 4 by the dashed lines.
%\begin{center}
%\fbox{Figure}
%\end{center}

Consequently, for a generic metric, the only Abelian instantons are flat
connections. On $S^{2} \times S^{2}$ there are no flat connections so
generically the moduli space is empty. We are also furnished with an
example of how one cannot
avoid a jump in the dimension of the moduli space along a one parameter
family of metrics on $X$ when $b_{2}^{+}=1$. $\lambda$ parameterizes the
relative volume of the two spheres. If one follows a path in the space
of metrics with $\lambda > 1$ to $ 1> \lambda$ then one cannot avoid
passing through $\lambda =1$ at which point there are solutions to self
dual equations.
%We summarise this as follows
%\begin{center}
%\fbox{Figure}
%\end{center}
This is a persistent problem for both Donaldson theory and Seiberg-Witten
theory.

The argument above breaks down if $b_{2}^{+}=0$ as then
$H^{2}_{-}(X,\RR)$ is the entire vector space (so, in particular, the
lattice lives there). For simply connected $X$, this can only
happen if $\chi = 2 - \sigma$. One
manifold for which this equality holds is $\overline{\CC \PP}^{2} $. In
this case $b_{2}=b_{2}^{-}= 1$.

\noindent\underline{ {\bf Moduli Space} }

The moduli space of Abelian instantons is taken to be the space of solutions
to the instanton equation modulo gauge transformations.
When Abelian instantons exist it is not difficult to describe the moduli
space. We may as well demand that $b_{2}^{+}(X)=0$, then any Abelian
instanton is described by
\be
F_{A} = 2\pi n_{i}[\Sigma_{i}] + da
\ee
where $d*da=0$ (as $d*F_{A}=0$). We also must fix the gauge, so we do
this by demanding $d*a=0$.

The restriction that $d*da=0$ implies
\be
\int_{X} a d*da = 0 \Rightarrow \int_{X} da * da = 0 \Rightarrow da=0 .
\ee
The last equality follows on noting that $||\o|| = \int \o *\o \geq 0$,
is a norm.
We now have the conditions that $da=d*a=0$ or that $a \in H^{1}(X,\RR)$.
The gauge invariant description of the points in the moduli space is to
consider $\int_{\ga_{i}}a$.

Actually this is not quite the end of the story as there are large gauge
transformations to take into account. Roughly these arise as follows:
a gauge transformation
can be thought of as a map $g: X \rightarrow U(1)$, these maps fall into
different classes because we can map the different 1-cycles non-trivially
into $U(1)$. For example, consider a non-trivial one cycle $\ga$ with
coordinates $0< \sigma \leq 2\pi$, then one has non-trivial maps
$g=\exp{(in \sigma )}$. Under such a gauge transformation $a \rightarrow a
+ g^{-1}id g $ one finds that $\int_{\ga}a \rightarrow \int_{\ga}a + 2\pi m$
so that these points of the moduli space are defined up to periodicity, they
lie on a torus. In general one has that the moduli space is indeed the
torus (the Jacobian of $X$)
\be
H^{1}(X,\RR)/H^{1}(X,\ZZ) .
\ee
This is the moduli space for fixed first Chern number.

\subsection{Topological Field Theory for Abelian Instantons}

We will now construct a topological model for Abelian instantons. Even
though in general they do not exist we will take the topology to be on
our side, namely, we will work with manifolds for which $b_{2}^{+}=0$.

The fields that we will need are: the gauge field $A_{\mu}$, its super
partner $\psi_{\mu}$ (Grassmann odd), and a scalar field $\f$ on the one
hand (these encode the geometry) while on the other hand one needs a self
dual tensor field $B^{+}_{\mu \nu}$, its super partner $\chi_{\mu \nu}$
(also self dual but Grassmann odd) and a pair of scalar super partners
$\fb$ and $\eta$ (the first Grassmann even the second odd). All the
fields are matched except $\f$ (but its superpartner is the ghost field
that one gets on gauge fixing $A_{\mu}$, and which I have suppressed).

The transformation rules are
\be
\begin{array}{ccc}
QA_{\mu} = \psi_{\mu} \, , & Q \psi_{\mu} = \partial_{\mu} \f \, , & Q\f = 0
\, , \\
Q \chi_{\mu \nu}^{+}= B_{\mu \nu}^{+} \, , & QB_{\mu \nu}^{+}=0 \, , &
\\
Q\fb = \eta \, , & Q\eta =0 \, . &
\end{array} \label{absym}
\ee
Notice that $Q^{2}= {\cal L}_{\f}$, where ${\cal L}_{\f}$ acts on the
fields as a gauge transformation. This means that, even though $Q^{2}
\neq 0$ acting on a gauge invariant function it will give zero. With
this in mind  we take the following as our action on any four-manifold $X$,
\bea
S &=& \{ Q , \int_{X} \chi^{+}F_{A}  +  \fb d*\psi \}
\nonumber \\
  & =& \int_{X} \left( B^{+} F_{A} - \chi^{+}d\psi + \eta d*\psi + \fb
d * d \f \right) .  \label{abact}
\eea

This action is inadequate because of the presence of zero modes. Clearly
there is one zero mode for each of $\f$, $\fb$, and $\eta$ (the constant
mode; $b_{0}=1$ on any manifold). There are also $b_{1}$ zero modes each
for $A$ and $\psi$, while there are $b^{+}_{2}$ zero modes for $B^{+}$ and
for $\chi^{+}$. We fix on a manifold with $b_{2}^{+}=0$, so that we need
not worry about zero for either $B_{+}$ or $\chi_{+}$. We can, by hand,
simply declare the zero modes of $\overline{\f}$ and $\eta$ to be zero.
One may also declare that the zero modes of $\f$ and the Faddeev-Popv
ghost field are also zero (though this is less natural).
One can do this in a BRST invariant manner \cite{bt}. We also do not
have to worry about the zero modes of the gauge field since they lie
naturally on a torus and integrating over them will give some finite
factor. We are left with the zero modes of $\psi$ to worry
about. However, we can soak these up by inserting operators of the form
$\int_{\ga} \psi$ into the path integral, where $\ga \in H_{1}(X, \RR)$.
Notice that these are BRST invariant operators. Everything is now more
or less under control-but what does it mean?

\noindent\underline{ {\bf Interpretation} }

Firstly the $B_{+}$ integral tells us that we are on the moduli space
\be
F_{A}^{+} =0 ,
\ee
(providing we also gauge fix). The $\chi_{+}$ integral enforces
\be
(1+*)d\psi =0 . \label{sdual}
\ee
This equation tells us that $\psi$ is tangent to the moduli space. To see
this we note that if $A$ is an abelian instanton and $A + \d A$ is also
an abelian instanton then $(F_{A}-F_{A+\d A})^{+}=0$ and consequently $(1+
*) d\d A = 0$, which is the equation satisfied by $\psi$.
{}From a geometrical point of view such a $\d A$ would correspond to a
tangent to the moduli space providing we also impose that $d*\d A =0$.
One really only wants
tangents which do not lie in the gauge directions and the
integral over $\eta$ imposes that $\psi$ has no components in that
direction, i.e.
\be
d*\psi =0 . \label{gf}
\ee

Indeed as $\psi$ is Grassmann odd it is most naturally thought of as a
one-form on the moduli space, and $Q$ then has the interpretation of
being the exterior derivative on ${\cal M}$.
How many such $\psi$ are there? We can answer this with a simple
`squaring' argument. From (\ref{sdual}), using the same argument as
above for the gauge field, we can conclude that
\be
0= \int_{X}d\psi(1-*)d\psi = \int_{X} d\psi *d\psi \Rightarrow d\psi=0.
\ee
Taken together with (\ref{gf}) this equation implies that $\psi$
is harmonic. We can expand $\psi = \lambda^{i} [\ga_{i}]$ for
$\lambda^{i}$ Grassmann parameters and $[\ga_{i}]$ a basis for
$H^{1}(X,\RR)$. The $\psi$ probe the tangent space to the torus and there
are as many of them as the dimension of the torus.

%\be
%\frac{1}{2\pi}\int_{\CC\PP^{1}}da = n .
%\ee
%Indeed as $\o$ spans $H^{2}(X,\ZZ)$ we may as well take $da/2\pi = \o$.

%The $A_{q}$ integral yields
%\be
%dB^{+}=0 \Rightarrow B^{+}=0 ,
%\ee
%as the constant mode in $B^{+}$ has already been eliminated. We should
%take stock. The only fields which have not been integrated out are the
%constant mode of $\f$ (we may as well denote it by $\f$) and $a$.

\noindent\underline{ {\bf Relationship with Maxwell Theory} }

To make some contact with Maxwell theory we note that we can write the
action (\ref{maxact}) as
\be
S= \frac{2}{g^{2}}\int_{X} F_{A}^{+} F_{A}^{+} + i\frac{\tau}{4\pi}
\int_{X} F_{A}F_{A} .
\ee
An equivalent theory is thus obtained on using
\be
i\frac{2}{g^{2}} \int_{X}B_{+}F_{A} + \frac{1}{2}\int_{X}B_{+}B_{+} +
 i\frac{\tau}{4\pi} \int_{X} F_{A}F_{A} \label{modact}
\ee
as the action.
There is a relationship between topological theories and physical theories
that comes about by `twisting'. This will be described in the next section,
but one part of the relationship that we need is that the actions of the
topological and physical theories are, at first, equivalent. This means that
what we really wish to consider as the action of the topological theory is
not just (\ref{abact}), but rather (\ref{modact}). This differs from the
bosonic part of (\ref{abact}) by a topological piece, the theta term, and
by a BRST trivial piece $B_{+}^{2}$,
\be
\int_{X} \{ Q , \chi_{+} B_{+} \} =  B_{+} B_{+} .
\ee
One can, therefore, use (\ref{modact}) as the bosonic part of the
topological action and maintain topological invariance. Our general
arguments tell us that, in principle, the addition of $B_{+}^{2}$ does not
change the results.

The upshot of this is that one calculates, within the
topological theory
\be
 \langle \exp{ \left( - i\frac{\tau}{4\pi} \int_{X} F_{A}F_{A}
\right) } \left( \prod_{i} \oint_{\ga_{i}} \psi \right)\rangle .
\ee
This automatically gives us
\be
Z(\tau)_{top} \sim \sum_{ n_{i} } \exp{ \left( -i \pi \tau n^{i}Q_{ij}n^{j}
\right) } .
\ee

%If one
%chases around the various determinants that arise on integrating out the
%fields then one would find that they all exactly cancel. This should come
%as no surprise as the supersymmetry guarantees this. We have, finally, the
%following finite dimensional integral to perform
%\bea
%Z& = & \sum_{n\in \ZZ} \int_{-\infty}^{+\infty}d\f \exp{( -s\f^{2} +
%it\f n )} \nonumber \\
% &= & \sum_{n\in \ZZ}\sqrt{ \frac{\pi}{ s} } \exp{\left(-\frac{n^{2}t^{
%2} }{4s} \right)}
%\eea
%which clearly depends on $t$ and $s$.

%We may put things on a more familiar footing by letting,
%\be
%s= \pi {\textstyle Im}\tau , \; \; \;
%\ee
%and $t^{2}= i ({\textstyle Im}\tau) \tau$.

%From the topological field theory point of view these choices are not
%natural. Rather, one would take
%\be
%S_{1} = \frac{1}{4\pi^{2}}\int_{\overline{\CC\PP}^{2}} \o \f F_{A}
%\ee
%The choice of the parameter in
%$S_{1}$
%is `canonical'. One fixes it in the following manner. The `observables'
%of the model come from various powers of
%\be
%\frac{1}{2\pi}\left(F_{A} + \psi + \f \right) = \frac{1}{2\pi}{\cal F}
%\ee
%which satisfies
%\be
%(Q-d){\cal F}=0.
%\ee
%The factor of $1/2\pi$ above comes from the normalization of the first
%chern class of $F_{A}$. ${\cal F}$ is known as the curvature on the
%universal bundle. We get $S_{1}$ by integrating $[{\cal F}/2\pi]^{2}$
%against the K\"{a}hler two-form $\o$. Likewise the parameter appearing
%in $S_{2}$ can be fixed by integrating $[{\cal F}/2\pi]^{2}$ against $\o
%\o$, but we leave it free for the moment.

%\begin{center}
%\fbox{To be continued}
%\end{center}

\section{Donaldson Theory}
Donaldsons original motivation for studying the moduli space of
instantons over a compact closed and simply connected four manifold $X$
was to get a handle on the possible differentiable structures that one
could place on $X$. The
dimension of the moduli space of instantons, for $SU(2)$, is
\bea
{\textstyle dim}{\cal M}&=& 8c_{2}-\frac{3}{2}(\chi + \sigma) \nonumber \\
& = & 8c_{2} - 3(1-b_{1}+b_{2}^{+}) .
\eea
For $c_{2}=1$ and $b_{1}=b_{2}^{+}=0$ the formal dimension is ${\textstyle
dim }{\cal M}=5 $. The space is depicted in figure 5.
\begin{center}
\fbox{Figure 5}
\end{center}

The sharp ends are the reducible connections. These are self-dual
connections for which the gauge group does not act freely. That is there
are non-trivial solutions to $d_{A}\f =0$. This happens precisely when
$A$ is an Abelian connection living, say, in the $3$ direction of
$su(2)$ and $\f$ constant also lying in the $3$ direction. We saw before
that we could avoid Abelian instantons {\em except} when $b_{2}^{+}=0$.
Of course not all self dual Abelian gauge fields are allowed, they must
satisfy
\bea
1 & =& \frac{-1}{8\pi^{2}}\int_{X} \tr F_{A}F_{A} \nonumber \\
&=& - \int_{X} \frac{F_{A}}{2\pi}\frac{F_{A}}{2\pi}
\eea
the notation being that in the first line one is dealing with the
$su(2)$ matrix
\be
\left( \begin{array}{cc}
F_{A} & 0 \\
0 & -F_{A}
\end{array} \right) .
\ee
How many solutions are there? The answer is $2b_{2}$. As
$H^{2}=H^{2}_{-}$ we have for a basis $[\Sigma_{i}]$ of $H^{2}(X, \ZZ)$
\be
\int_{X} [\Sigma_{i}]*[\Sigma_{j}] = G_{ij}
\ee
but to simplify life a little suppose $G_{ij}= \d_{ij}$. As the
$[\S_{i}]$ are all anti self-dual $*[\Sigma_{i}]=-[\Sigma_{i}]$ we
obtain
\be
\int_{X} [\Sigma_{i}][\Sigma_{j}] = -\d_{ij} .
\ee
One expands $F_{A} = 2\pi n_{i}[\Sigma^{i}]$ and so the constraint
becomes $n_{i}n^{i}=1$. All vectors of the form $(0,\dots,\pm 1,
\dots,0)$ satisfy this and in a vector space of dimension $b_{2}$ there
are $2b_{2}$ such vectors. When $G_{ij} \neq \d_{ij}$ one comes to the
same counting, by diagonalising $G$ (recall that it is a symmetric and
non-degenerate matrix).

It is possible to show that the conical singularities, look like cones
on $\CC\PP^{2}$ and so one can `smooth' things out by replacing the
singularities with copies of $\CC\PP^{2}$.

The other end of the moduli space is a copy of the original manifold
$X$. Recall that $SU(2)$ instantons on $\RR^{4}$ (or $S^{4}$) are
parameterized by their position and their scale. This is also true on a
compact manifold (thanks to some work of Taubes). When one shrinks the
scale down to zero (almost) they are parameterized only by their
position on $X$. Saying that backwards: $X$ parameterizes the zero size
instantons and hence appears at the end of the moduli space.

It takes a lot of analysis, but one can show that the space one finally
obtains is a smooth, orientable five manifold with a boundary, on one
side that is $X$ and on the other, is made up of copies of
$\CC\PP^{2}$ thus giving one a useful `cobordism'. Using this
relationship between the various spaces, one can, for example, prove
that if $Q \neq 0$ and $Q(\S,\S) \neq 1, \forall [\S]  \in H^{2}(X,\ZZ)$
then $X$ is not smoothable!

All of this was the original motivation. Later Donaldson realized that
one could work with all sorts of instanton moduli spaces (of various
dimensions). One could then  define
cohomology classes on those spaces which, under good conditions, would
be `topological' invariants that one could associate with the underlying
manifold $X$.

To see what these invariants distinguish we go back to the
intersection form $Q$. For a simply connected, connected, closed, compact
and smooth
four-manifold X, $Q$, determines the signature $\tau(X)$ (as the signature
of the matrix), the Euler character $\chi(X)$, wether $X$ is a spin
manifold (as we saw in our study of Maxwell theory) and even if $X$
admits an almost complex structure! In a sense $Q$, encodes all of the
classical invariants of
a manifold. The topological question of relevance is: If someone hands
you a $Q$ have they given you a particular four-manifold? The answer is
an emphatic no! There are many different four-manifolds all sharing the
same $Q$. The Donaldson invariants, or cohomology classes, can
distinguish between manifolds which share the same intersection form and
so are nothing like any of the classical invariants.

It turns out that Wittens topological field theory gives a
ready description of these classes and so we turn to that.

\subsection{Topological Field Theory of ${\cal M}$}

I will be very brief in my description here. A proper job would
require us to go into details that are not crucial to the present
discussion and so I would refer the reader to \cite{bbrt} for a more
detailed account.

In order to write down an action that devolves to an integral over the
moduli space of instantons we adopt the same field content as in the
$U(1)$ case, except the fields all take values in the adjoint
representation. In a nutshell
\be
\begin{array}{lcl}
QA = \psi & & Q\psi = d_{A}\f \\
Q\chi_{+}=B_{+} & & QB_{+} = [\chi_{+}, \f ] \\
Q\overline{\f}= \eta & & Q\eta = [\overline{\f} , \f ] \\
 & Q\f =0 &
\end{array}
\ee
with $Q^{2}= {\cal L}_{\f}$. We can now write down the action
\bea
S&=& \int_{X} \{ Q, \left( \chi_{+}F_{A} + \frac{s}{2}\chi_{+}B_{+} +
t\overline{\f}d_{A}*\psi  \right) + u \eta * [\overline{\f},f] \} \nonumber \\
 &=& \int_{X}\left( B_{+}F_{+} - \chi_{+}d_{A}\psi +
\frac{s}{2}B_{+}B_{+} -\frac{s}{2}\chi_{+}[\chi_{+}, \f]  + t\eta d_{A} *
\psi   \right. \nonumber \\
 & & \left. \; \; + t\overline{\f} d_{A}*d_{A} \f + t\overline{\f}\{ \psi
, *\psi \}  + u [\overline{\f}, \f] * [\overline{f}, \f] - u \eta * [\eta ,
\f]
\right). \label{donact}
\eea

When one takes $s=0$, we have delta function support on the moduli-space
of instantons. It can happen that one needs to thicken things out, and
keep $s \neq 0$ until the end of the calculation, especially if there
are $B_{+}$ zero modes. However, as long as everything remains well
defined, we can vary the parameters $s$, $t$ and $u$ at will.

\subsection{Observables}

The observables of the theory can be obtained by the descent equation
\be
(Q-d)\Tr \left( F_{A} + \psi + \f \right)^{n} = 0 . \label{descent}
\ee
Keeping track of the degree of the forms and their Grassmann numbers
allows us to derive some useful identities. For example
\be
Q\Tr \f^{2}(x) =0
\ee
tells us that ${\cal O}^{0}(x)=\Tr \f^{2}(x)$ is a good observable (BRST
invariant), while
\be
d\Tr\f^{2}(x) = \{ Q , \Tr \psi(x) \f(x) \} \label{id1}
\ee
tells us that in BRST invariant correlation functions, one can prove
that the correlation function does not depend on the point $x$ at which
on places $\Tr \f^{2}$, as behoves a topological observable!

We can construct another set of BRST invariant states by integrating
$\Tr  \psi \f$
over one-cycles $\ga_{i} \in H_{1}(X, \RR)$ (as one can check by using
(\ref{id1}),
\be
Q\Tr\int_{\ga_{i}} \psi \f =0 .
\ee
In order to establish that these observables depend on on the homology
class and not on the representative we need yet another useful identity
namely
\be
d \Tr \psi \f  = \{ Q, \Tr \left( F_{A}\f + \psi\psi \right) \} .
\label{id2}
\ee
Using this one has
\bea
\Tr \int_{\ga_{i} + \partial \sigma_{i}} \psi \f - \Tr \int_{\ga_{i}}
\psi \f & =& \int_{\partial \sigma_{i}}\Tr (\psi \f) \nonumber \\
 & =& \int_{\sigma_{i}} d \Tr (\psi \f) \nonumber \\
&=& \{ Q , \int_{\sigma_{i}} \Tr \left( F_{A}\f + \psi \psi \right) \} .
\eea
Hence ${\cal O}^{1}(\ga_{a})= \int_{\ga_{a}}\Tr (\psi \f)$ is a
topological observable.

One can show that
\be
{\cal O}^{2}(\S_{i})= \int_{\S_{i}}\Tr \left(F_{A}\f + \psi \psi \right)
\ee
is BRST invariant, and I leave it as an exercise to show that it only
depends on the class of $\S_{i} \in H_{2}(X,\ZZ)$.

The observables, ${\cal O}^{i}$, we have thus constructed can be thought
of as $(4-i)$ forms on the moduli space. This interpretation is natural,
since $\psi$ is to be thought of as a one-form on ${\cal M}$ and $Q$ is
exterior differentiation, so a glance at the transformation rules will
convince the reader that $\f$ must be a two-form on the moduli space.
The expectation value of products of the observables will therefore
correspond to the integration over ${\cal M}$ of products of elements of
the $(4-i)$ cohomology classes. This means that what we would be doing, if
we could do the calculations, is
intersection theory on the moduli space.

The formalism has been around for some time, but nobody had managed to
perform any calculations with it. Witten \cite{witsym}, turned the
calculation of the Donaldson polynomials into a problem of calculating
the correlation functions of the ${\cal O}^{i}$ in $N=1$ supersymmetric
Yang-Mills theory. By exploiting various features of the physical theory
Witten was able to get formulae for the invariants on any K\"{a}hler
four manifold. Turning the argument around, the purported correspondence
between the physical and topological theories puts constraints on the
physical theory! I will not review this work here, however I recommend
the reading of it as it stands as the isolated example of the evaluation
of the Donaldson
invariants from the physics point of view.

In order to make contact with
the monopole equation invariants let me just note some formulae for
manifolds $X$ of simple
type. Let
\be
D(\la , v) = \langle \exp{ \left( \la {\cal O}^{0}(x) + \sum_{i}\a_{i}
{\cal O}^{2}(\S_{i})
\right) } \rangle ,
\ee
where $v = \sum_{i} \a_{i} [\S_{i}]$ with $\la$ and $\a_{i}$ complex
numbers. Recall the notation $v^{2}= \sum_{i,j} \a_{i} \a_{j}
Q(\S_{i},\S_{j})$ and for any $x \in H^{2}(X, \ZZ)$, $v.x = \sum_{i}
\a_{i} Q(x,\S_{i})$. A manifold $X$ is said to be of simple type, if
\be
\frac{\partial^{2}D(\la , v )}{\partial \la^{2}} = 4D(\la , v) .
\ee
A theorem, due to Kronheimer and Mrowka \cite{km}, states that for
manifolds of simple type
\be
D(\la , v) = \exp{ \left( \frac{v^{2}}{2}  + 2 \la \right)} \sum_{x}
n_{x} e^{v.x}
\ee

What the Seiberg-Witten invariants will correspond to are the unknowns
in this formula, namely the $n_{x}$. Rather than pursuing this now, our
attention is
directed to the original derivation of the topological theory from $N=2$
supersymmetric Yang-Mills theory.

%\begin{center}
%\fbox{To be continued}
%\end{center}

\subsection{Relationship to $N=2$ Super Yang-Mills Theory}
To establish the relationship between the topological theory and a
physical theory one needs the notion of twisting. It is easiest to start
with the $N=2$ theory. From various lectures we have seen that the
theory is described by one $N=2$ chiral superfield with components
\be
\begin{array}{c}
  A_{\mu} \\
\begin{array}{lr}
 \lambda^{1} & \lambda^{2}
\end{array} \\
 \varphi
\end{array}
\ee
The action for this theory is
\bea
S_{N=2}& =& \frac{1}{g^{2}} \int_{\RR^{4}}d^{4}x \, \Tr
\left(-\frac{1}{4}F_{\mu \nu}F^{\mu \nu} -i \overline{\la}^{\dot{\a}}_{i}
\sigma^{\mu}_{\a \dot{\a}} D_{\mu}\la^{\a i} -D_{\mu}\overline{\varphi}
D_{\mu} \varphi \right. \nonumber \\
& & \; \; \; \left. -\frac{1}{2}[\varphi, \overline{\varphi}]^{2}
-\frac{i}{\sqrt{2}}\overline{\varphi}\eps_{ij}[\la^{\a i},\la_{\a}^{j}]
+ \frac{1}{\sqrt{2}} \varphi \eps^{ij} [\overline{\la}_{\dot{a} i},
\overline{\la}^{\dot{\a}}_{j}] \right) \label{supaction}
\eea

The $N=2$ theory has for global symmetry the Lorentz group
$SU(2)_{L}\times SU(2)_{R} $  as well as an internal $SU(2)_{I}$ which
acts on $\lambda^{i} = (\lambda^{1}, \lambda^{2})$, thus exchanging the
two supersymmetries. The quantum numbers of the fields under
$ SU(2)_{L}\times SU(2)_{R} \times SU(2)_{I} $ are
\bea
A_{\mu} &:& \left(\frac{1}{2}, \frac{1}{2}, 0 \right) \nonumber \\
\lambda^{1} &:& \left( \frac{1}{2}, 0 , \frac{1}{2} \right) \nonumber \\
\lambda^{2} &:& \left( 0, \frac{1}{2},  \frac{1}{2} \right) \nonumber \\
\varphi &:& \left( 0,0,0 \right)
\eea

Twisting amounts to redefining the Lorentz group to be $SU(2)_{L} \times
SU(2)_{R'}$ where $SU(2)_{R'}$ is the diagonal sum of $SU(2)_{R}$ and
$SU(2)_{I}$. The transformation of the fields under $SU(2)_{L} \times
SU(2)_{R'}$ are
\bea
A_{\mu} &:& \left(\frac{1}{2}, \frac{1}{2} \right) \nonumber \\
\lambda^{1} &:& \left( \frac{1}{2},  \frac{1}{2} \right) \nonumber \\
\lambda^{2} &:& \left( 0, 1 \right) \oplus \left( 0, 0 \right) \nonumber \\
\varphi &:& \left( 0,0 \right) .
\eea
We are now in a position to match these fields with those in the
topological theory. The gauge field is of course the same in either
theory as is $\varphi = (\f , \overline{\f})$. $\lambda^{1}$ is now a
vector and is what we called $\psi_{\mu}$ in the topological theory.
$\lambda^{2 }$ is now a sum of a scalar and a self dual two-form and
hence corresponds to $(\chi_{+}, \eta)$. Notice that nothing in sight
can be identified with $B_{+}$; this is no cause for alarm as $B_{+}$ is a
multiplier field and may be eliminated algebraically.

One can also determine the new weights of the supersymmetry charges.
Originally these were
\bea
Q_{\a}^{I} &:& \left( \frac{1}{2}, 0 , \frac{1}{2} \right) \nonumber \\
\overline{Q}_{\dot{\a}}^{I} &:& \left( 0, \frac{1}{2}, \frac{1}{2}
\right) ,
\eea
and after the twisting become $(1/2,1/2)$ and $(0,0) \oplus (0,1)$
respectively. The
$(0,0)$ component is a scalar supersymmetry charge and is what we have
been calling $Q$. The charges may be denoted by $Q_{\mu}$, $Q_{\mu
\nu}^{+}$ and $Q$.

%\begin{center}
%\fbox{To be continued}
%\end{center}

A comparison of (\ref{donact}) and (\ref{supaction}) with the
identifications of the fields that we have made shows that all of the
terms in (\ref{supaction}), after twisting, generate the elements of the
action (\ref{donact}).

\noindent\underline{Remarks}

\begin{itemize}
\item The topological action (\ref{donact}) with arbitrary parameters $s$
and $t$ is only guaranteed to be invariant under $Q$. For special values of
these parameters the theory may enjoy more symmetry.
\item One can twist the $N=2$ action and this will correspond to
(\ref{donact}), for certain values of $s$ and $t$, up to theta terms.
\item The $N=2$ action in twisted from will be invariant under
$Q_{\mu}$, $Q_{\mu \nu}^{+}$ and $Q$.
\end{itemize}

\subsection{Relationship with the Monopole Equations}
Donaldson theory, as described above, is related to $N=2$ super
Yang-Mills  theory. Indeed the usual way to relate the
correlation functions in the field theory to the Donaldson invariants is
at weak coupling in the ultraviolet. This corresponds to $u \sim \infty$
in the quantum moduli space, as explained in the lectures of
Alvarez-Gaume. The beauty of the topological theory is that it is
coupling constant independent and so if one could also evaluate the
physical theory at strong coupling (in the infrared) one would get a
different, though, equivalent, description of the Donaldson invariants.

Seiberg and Witten \cite{sw} found that the infrared limit of the $N=2$
theory in
the infrared is equivalent to a weakly coupled theory of Abelian gauge
fields coupled to `monopoles'. In the $u$ plane this region corresponds
to the vicinity of the points $|u|=1$. Away from these points only the
Abelian gauge field is massless while at the degeneration points $u =
\pm 1$, the monopoles also become massless.

As the theory is weakly
coupled at $|u|=1$, one is tempted to twist the physical theory at those
points to get a different description of the Donaldson polynomials. At
those points there is the photon plus an $N=2$ hypermultiplet (also
called a scalar multiplet) of two Weyl fermions, $q$ and
$\tilde{q}^{\dagger}$ and complex bosons,
massless `monopoles', $B$ and $\tilde{B}^{\dagger}$. We put them into a
diamond:
\be
\begin{array}{c}
  q \\
\begin{array}{lr}
 B & \tilde{B}^{\dagger}
\end{array} \\
  \tilde{q}^{\dagger}
\end{array} .
\ee

Once more the $SU(2)_{I}$ symmetry acts on the rows and thus
non-trivially only on ($B$, $\tilde{B}^{\dagger}$). If we twist, $q$ and
$\tilde{q}^{\dagger}$ remain Weyl spinors, however, $B$ and
$\tilde{B}^{\dagger}$ transform as $(0,1/2)$, that is, as spinors.

The theory, before twisting, is made up of the Abelian version of
(\ref{supaction}) plus the $N=2$ matter action for the fields $(q,
\tilde{q}^{\dagger}, B, \tilde{B}^{\dagger})$. A part of the bosonic
matter action is
\be
\int_{\RR^{4}} \left( \overline{D_{\mu}B_{i}}D^{\mu}B^{i} +
\frac{1}{2}(\overline{B_{i}} B^{i})^{2} + \dots \right) .
\ee
To write down the corresponding topological action, we just need to
twist the supersymmetric action which leads to
\bea
& & \int_{X} \sqrt{g} \left(  \frac{1}{2} F_{+}^{2} + g^{\mu \nu}D_{\mu}
 \overline{M}.D_{\nu}M +
\frac{1}{2} (\overline{M}M)^{2} + \frac{R}{4}\overline{M}M + \dots
\right) \nonumber \\
& & \equiv \int_{X} \left( \frac{1}{2}(F_{\mu \nu}^{+} +
\frac{i}{2}\overline{M}
\sigma_{\mu \nu}M)^{2} + \slash{D} M \slash{D} \overline{M} + \dots
\right) . \label{twistedmon}
\eea
The notation will be explained in the next section, and the equivalence
will follow from the Weitzenboch formula derived presently. Note that
$M$ is the spinor that $B$ has been twisted to.

This is one way to
do it, but we could also start from the monopole equations, the absolute
minima of the action above (see the next section), and apply the
formalism developed at the start of these lectures and follow our noses.
This is a very straightforward exercise, which I shall not do, if you
try it and get stuck see \cite{ruibin}

%\begin{center}
%\fbox{To be continued}
%\end{center}
The above discussion needs to be re-assessed on a compact manifold.
Though heuristically correct a more detailed analysis is required to
show that on a compact manifold the monopole equations arise from the
twisted $N=2$ theory. Arguments in favour of this have been presented by
Witten \cite{witsw}. Regardless of the derivation one can simply take
the monopole equations as given and then ask what implications they have
for the study of four-manifolds.

\section{Seiberg-Witten Theory}
This section is longish as I have tried to give a rather complete
treatment of the various pieces that go into the analysis of the
Seiberg-Witten equations. As we go along I will make some comments on
the dimensionally reduced version of the equations in footnotes. The
interest in the lower dimensional case stems from the fact that the
dimensionally reduced Donaldson theory gives topological invariants for
lower dimensional manifolds. The one of prime interest is three
dimensions, where the equivalent of the Donaldson theory calculates the
Casson invariant. One would like to know what the relationship between
the (reduction of the) equations that we are about to study and the
Casson invariant, if any, is.

The equations\footnote{In three dimensions the first equation reads
$F_{\mu \nu}= \frac{1}{2}\eps_{\mu \nu \la}\overline{M}\ga^{\la}M$,
while in two dimensions a natural analogue is $F_{\mu \nu}= \frac{1}{2}
\eps_{\mu \nu}\overline{M}M$.} that we will be studying in this section are
\bea
F_{\mu \nu}^{+}&=& -\frac{i}{2}  \ol{M} \sigma_{\mu \nu} M \, , \nonumber
\\
\ga^{\mu}D_{\mu} M &=& 0 \, \label{sw}
\eea
where $M^{\a}$ is a Weyl spinor, satisfying $\ga_{5}M=M$, $\ol{M}_{\a}$
its complex conjugate. The gamma matrices satisfy $\{ \ga_{\mu},
\ga_{\nu} \} = 2 g_{\mu \nu}$ and $\sigma_{\mu\nu }= \frac{1}{2}[\ga_{\mu}
, \ga_{\nu}]$. In order to define the covariant derivative of a spinor on
an arbitrary 4-manifold we need to introduce a spin connection
$\o_{\mu}^{ab}$ with this in hand one has
\be
D_{\mu} = \partial_{\mu} +iA_{\mu} + \frac{1}{4}\o_{\mu}^{ab}
\sigma_{ab} \, .
\ee

A complete set of conventions for $\ga$-matrices, spinors, the spin
connection and  vierbeins is given in the appendix.

There are a few points that we ought to check about these equations.
Firstly the i factor on the right hand side of the first equation in
(\ref{sw}) is needed as the gauge field $A_{\mu}$ is taken to be real.
Secondly, as the self dual part of $F_{\mu \nu}$ appears on the left
hand side, only the self dual part should appear on the right hand side.
This is indeed the case. With our definitions
\be
\sigma_{\mu \nu } \ga_{5} = \frac{1}{2} \eps_{\mu \nu}^{\; \; \; \; \kappa
\lambda}\sigma_{\kappa \lambda}
\ee
so that,
\bea
\ol{M}\sigma_{\mu \nu}M &=& \frac{1}{2} \eps_{\mu \nu}^{\; \; \; \; \kappa
\lambda}\ol{M}\sigma_{\kappa \lambda}M \nonumber
\\
&=& \frac{1}{2}\left( \d^{\kappa}_{\mu}\d^{\lambda}_{\nu} +
\frac{1}{2} \eps_{\mu \nu}^{\; \; \; \; \kappa
\lambda}\right) \ol{M}\sigma_{\kappa \lambda}M
\eea
which is self dual\footnote{Had we changed the sign of $\ga_{5}$ we
would have found this combination to be anti-self dual}.

%One must also check that the Bianchi identity on the field strength does
%not impose additional constraints. For self dual connections the Bianchi
%identity can be cast as $g^{\mu \nu}\partial_{\mu}F_{\nu
%\lambda}^{+}=0$. In this form

The sign on the
right hand side of the first equation in (\ref{sw}) is important as well.
We will see, shortly, that with this choice of sign strong restrictions
can be placed on the solution set\footnote{One can flip the sign if one
takes $M$ to have opposite charge, ie. $M \in \Gamma(S_{+}\otimes L^{-1})$.}.

Our main
objective is to analyse these equations in more detail and to see what
they imply for 4-manifolds, but first a digression.

\subsection{Spin and Spin$_{\CC}$ Structures}
There is a (would be) catch to writing down these equations. The bad news
is that on many manifolds there are topological obstructions to defining
spinors. The good news is that spinors can be defined on any smooth
compact orientable 4-manifold if they are coupled to gauge fields that
satisfy a certain restriction. We will firstly review the obstructions
and then proceed to the coupling to gauge fields.

To see that there is indeed a problem, consider the index of the Dirac
operator on $X$. The index counts the number of left-handed
minus the number of right-handed solutions to the Dirac equation, and is
therefore an integer. The Atiyah-Hirzebruch-Singer index theorem gives us a
topological formula for the index, namely
\be
{\rm Index} \ga^{\mu}\nabla_{\mu} = - \frac{\tau(X)}{8}.
\ee
If the right hand side is to be an integer then the signature of $X$
must be divisible by $8$. If it is not divisible by $8$ then the
assumption that we had spinors in the first place is in fact not correct.
The standard example of a manifold that does not admit spinors is
$\CC\PP^{2}$ \cite{hp}. The second Betti number for $\CC\PP^{2}$ is
$b_{2}=b_{2}^{+} =1 $ and hence $\tau(\CC\PP^{2})= 1$, so that
$\CC\PP^{2}$ is not spin. This conclusion is consistent with the theorem
that for a simply connected $X$ the intersection form must be even for
$X$ to admit a spin structure.

A way of getting around the problem is to introduce a
spin$_{\CC}$ structure. This is nothing mysterious, it means that we
couple a $U(1)$ gauge field to the spinor, that is, we consider now
charged spinors. The index in this case
reads
\be
{\rm Index} ( \ga^{\mu}D_{\mu}) = -\frac{\tau(X)}{8} +
\frac{1}{2}c_{1}(L)^{2}.
\ee

Suppose that we choose the line-bundle $L$ so that the extra term
cancels the offending fraction. For example, on $\CC\PP^{2}$ we take
\be
c_{1}(L)= \frac{F_{A}}{2\pi}= (n+ \frac{1}{2})\o
\ee
so that
\be
\frac{1}{2}c_{1}(L)^{2}= \frac{n}{2}(n+1) + \frac{1}{8}.
\ee
Now the index of the Dirac operator is an integer
\be
{\rm Index}(\ga^{\mu}D_{\mu}) = \frac{n}{2}(n+1) .
\ee

Of course such $U(1)$ bundles are not defined, but then neither are the
spinor bundles. What we have here is a coherent cancellation of the
obstructions to defining both bundles so that the product makes sense.
Let me reiterate this, $S_{+}$ and $L$ do not make sense as bundles,
however, the product $S_{+} \otimes L$ is an eminently reasonable
bundle to consider. There are different possibilities for making the
product well defined, so that, in principle we should sum over all
possible spin$_{\CC}$ structures. However, as we will sum over all
possible $U(1)$ bundles we are essentially summing over the possible
spin$_{\CC}$ structures. From now on, when $X$ is not spin we understand
that it is equipped with a spin$_{\CC}$ structure.

For the reader who would like to know more about the obstructions to
spinors and the existence of spin$_{\CC}$ I would recommend the reading
of \cite{gsw} chapter 12, \cite{hp} and for the more mathematically
minded \cite{lm}, in that order.

%\begin{center}
%\fbox{To be continued}
%\end{center}

%\subsection{$U(1)$ Bundles}
%\begin{center}
%\fbox{To be continued}
%\end{center}

\subsection{The Invariant}

\noindent\underline{The Dimension of ${\cal M}$}

\noindent Let $(A,M)$ be a solution to the monopole equations. We want
to use an index calculation to estimate the dimension of the moduli
space of solutions. For $(A+\d A , M + \d M)$ to be a nearby solution we
require
\be
(d\d A)^{+} =  -\frac{i}{2} \d \overline{M}.\sigma . M -\frac{i}{2}
\overline{M}.\sigma .\d M
\ee
and
\be
\ga^{\mu}D_{\mu} \d M + \ga^{\mu}\d A_{\mu} M =0
\ee
to hold. The number of linearly independent $(\d A, \d M)$ tells us the
dimension
of the moduli space. It is often possible to show that the answer for the
index of an operator does not depend on the coefficients that appear in
it. In our case this means that we will search for solutions around
$M=0$. (We cannot set $A=0$ as that will change the bundle we are
sitting on). So we wish to count solutions to
\be
(d\d A)^{+} = 0 \, , \; \; d*\d A =0 \, , \; \; \ga^{\mu}D_{\mu} \d M =0 \, .
\ee
Notice that we have gauge fixed.

It is useful to define an elliptic operator $T$ by
\bea
& & T : \Omega^{1}(X) \oplus (S^{+} \otimes L) \rightarrow \Omega^{0}(X)
\oplus \Omega_{+}^{2}(X) \oplus (S^{-} \otimes L) \nonumber \\
& & T:(\d A , \d M) \mapsto (*d*\d A , (d \d A)^{+}, \slash{D} \d M) \,
. \label{Top}
\eea
We calculate the index of $T$ (this is insensitive to us setting $M=0$).
This is straightforward to do, for $T$ splits up into the sum of the index
of the operator $d + *d*$, acting on self dual two forms, and on the Dirac
operator from $S^{+} \otimes L$ to $S^{-} \otimes L$.

The virtual dimension is, therefore
\be
d = -\frac{2\chi(X) + 3 \tau(X)}{4} + c_{1}(L)^{2} \, .
\ee
When $0>d$ there are generically no solutions to the monopole equations.
More interesting for us is when $d=0$. Let $x= -c_{1}(L^{2})=
-2c_{1}(L)$, then $d$ vanishes precisely when
\be
x^{2}= 2\chi(X) + 3\tau(X) \, . \label{dim0}
\ee
Generically, when $x$ satisfies $x^{2}= 2 \chi(X) + 3\tau(X)$, there will be
a set of $t_{x}$ isolated solutions to the monopole equations (up to gauge
transformations). Label these points by $P_{i,x}$, $i=1 , \dots t_{x}$.
We can now define

\vspace{.6cm}

\noindent \underline{ The Seiberg-Witten Invariant}

Fix an $x$ that satisfies (\ref{dim0})
and to each $P_{i,x}$ associate a sign $\eps_{i,x}= \pm 1$- the sign of
the determinant of $T$.
Our discussion of the Euler character of a Riemann surface has prepared
us for the following definition. The Witten invariant, $n_{x}$, is the
integer
\be
n_{x} = \sum_{i} \eps_{i,x} \, . \label{inv}
\ee

%Under certain conditions

\vspace{.8cm}

\noindent\underline{Perturbations}

We have seen that the formal dimension of the moduli space is
\be
d = \frac{1}{4}\left( 2\chi(X) + 3 \tau(X) \right) +
c_{1}(L). c_{1}(L) \, . \label{dim}
\ee
In these notes we will be interested in the case where $d=0$. However,
this is the vanishing of the formal dimension. What we would really like
is to have that the formal and true dimension of ${\cal M}$
coincide. To achieve this one may have to perturb the equations. We do
this by passing to
\bea
F_{\mu \nu}^{+} & =& -\frac{i}{2}\overline{M}\sigma_{\mu \nu}M + p_{\mu
\nu } \, , \nonumber \\
\slash{D}M &=& 0 \, , \label{psw}
\eea
with $p$ some real self-dual two form. Here is a nice fact due to
Taubes.

\noindent \underline{{\bf Fact \cite{taubes1}}}: Let $X$ be a compact,
oriented,
4-manifold with $b_{2}^{+} \ge 1$ and with a symplectic two form $\o$,
then the space of solutions to (\ref{psw}), ${\cal M}(p)$, will be a
smooth manifold for a generic choice of $p$ with dimension (\ref{dim}).

In this situation we are in the same position as we were for the
calculation of the Euler character previously. When $d=0$, for a
judicious, though generic, $p$, and with $b_{2}^{+}>1$, ${\cal M}(p)$
is a finite union of signed points and the Witten invariant is the sum
over these points of the corresponding $\pm 1$'s.

When we come to considering K\"{a}hler surfaces one can be very explicit
about the perturbation. Indeed, following Witten, we will give a thorough
description of the perturbed moduli space.

\subsection{Bochner-Kodaira-Lichnerowicz-Weitzenb\"{o}ck Formula}
Given a set of first order equations, like the monopole equations, there
is a technique for extracting some very useful information. The idea
goes back to Weitzenb\"{o}ck, but was used most effectively by Bochner. In
the context of the Dirac equation, it was Lichnerowicz who first derived
that there are no harmonic spinors on a compact closed spin manifold
whose scalar curvature is positive definite, using this technique \cite{l}.
In the complex domain, Kodaira has also put this idea to good use.
For reasons that will become apparent shortly, I will simply refer to this
as the squaring argument.
Before getting to that we need an identity, namely (see the appendix for
some more details)
\bea
\slash{D}_{A}^{2} &= & \gamma^{\mu}\gamma^{\nu}D_{\mu}D_{\nu} \nonumber \\
&=& \left(\frac{1}{2}\{ \gamma^{\mu}, \gamma^{\nu}\} +
\frac{1}{2}[\gamma^{\mu},\gamma^{\nu}]\right)D_{\mu}D_{\nu}
\nonumber \\
&=& D^{\mu}D_{\mu} + \frac{1}{2} \sigma^{\mu \nu} [D_{\mu}, D_{\nu}]
\nonumber \\
&=& D^{\mu}D_{\mu} + \frac{1}{2} \sigma^{\mu \nu} F_{\mu \nu} -
\frac{R}{4} .
\eea
This formula holds irrespective of the dimension of the space that we
are working on.

Now onto the squaring argument. Let
\be
s_{\mu \nu}= F^{+}_{\mu \nu} + \frac{i}{2}  \ol{M} \sigma_{\mu \nu} M \, \;
\;  \; k^{\a}= \left(\slash{D} M \right)^{\a} \, ,
\ee
and using the above formula we have
\be
 \int_{X}d^{4}x \sqrt{g}\left(\frac{1}{2}|s|^{2}+|k|^{2}\right)\equiv
\int_{X} d^{4}x\sqrt{g}\left( \frac{1}{2}g^{\mu \kappa}g^{\n \lambda}s_{\mu
\nu} s_{\kappa \lambda} + \ol{k}_{\a}k^{\a} \right) =0 \, .
\ee
The aim is to cast this identity into a useful form. In order to do this
we look at the separate parts that appear and simplify them.

Notice that
\bea
\int_{X}d^{4}x\sqrt{g} \ol{\slash{D}M}\slash{D}M &=& -\int_{X}d^{4}x
\sqrt{g} \ol{M}
\slash{D} \slash{D}M \nonumber \\
&=& - \int_{X}d^{4}x\sqrt{g}\ol{M} \left( D^{\mu}D_{\mu} +
\frac{i}{2}\sigma^{ \mu
 \nu} F_{\mu \nu }-\frac{R}{4}\right) M \nonumber \\
&=& - \int_{X}d^{4}x\sqrt{g}\ol{M} \left( D^{\mu}D_{\mu} +
\frac{i}{2}\sigma^{\mu  \nu} F_{\mu \nu }^{+}-\frac{R}{4}\right) M \, ,
\eea
the last line follows as we know the combination $\ol{M}\sigma M$ is self
dual.

One more relationship that we need is by way of a Fierz identity,
\bea
\d^{\a}_{\b}\d^{\ga}_{\eps}&=& \frac{1}{4}\left(
\d^{\a}_{\eps}\d^{\ga}_{\b} +
(\ga_{\mu})^{\a}_{\; \eps}(\ga^{\mu})^{\ga}_{\; \b} +
(\ga_{5})^{\a}_{\; \eps}(\ga_{5})^{\ga}_{\; \b} \right. \nonumber \\
& & \; \; \; \; \; \left. - \frac{1}{2}
(\sigma_{\mu \nu})^{\a}_{\; \eps}(\sigma^{\mu \nu})^{\ga}_{\; \b}
-(\ga_{\mu}\ga_{5})^{\a}_{\; \eps}(\ga^{\mu}\ga_{5})^{\ga}_{\; \b}\right) \, .
\eea
Multiply this equation with $M_{\a}M_{\ga} \overline{M}^{\b}
\overline{M}^{\eps }$ and recall that the spinor $M$ is Weyl to obtain
\be
- \frac{1}{8}\overline{M}\sigma_{\mu \nu}M \overline{M}\sigma^{\mu \nu}M =
\frac{1}{2} \left(\overline{M}M\right)^{2} \, . \label{fierz}
\ee

Putting all the pieces together, one arrives at\footnote{The formula is
almost the same for the two and three dimensional theories. One replaces
$F^{+}$ with $F$ and rather than $\frac{1}{2}|M|^{4}$ one has
$\frac{1}{4}|M|^{4}$. Some of the easier consequences of the vanishing
theorems also follow in these cases.}
\bea
& &  \int_{X}d^{4}x\sqrt{g}\left( \frac{1}{2} |s|^{2} + |k|^{2} \right)
\nonumber \\
& & = \int_{X}d^{4}x \sqrt{g}\left( \frac{1}{2} |F^{+}|^{2} + g^{\mu
\nu}\overline{D_{\mu}M}. D_{\nu}M + \frac{1}{2} |M |^{4} +\frac{1}{4}R
|M|^{2} \right) \, . \label{blw}
\eea
Notice that the cross terms, $F^{+}\overline{M}\sigma M$, which are
present in both $|s|^{2}$ and $|k|^{2}$ cancel in the sum. This is why
the sign in the first of (\ref{sw}) is important.

One recognizes (\ref{blw}) as part of the twisted monopole action
(\ref{twistedmon}). As it is the sum of squares, the absolute minimum of
the twisted monopole action will be where the monopole equations are
satisfied. It is no surprise then that the topological theory that one
gets on twisting the $N=2$ model in the infrared is the ``right'' one.

\subsection{Vanishing Theorems}
The vanishing of (\ref{blw}) puts some constraints on the solution set
of (\ref{sw}). For example, if there is a metric on $X$ for which $R> 0$
then all the terms in (\ref{blw}) are positive and so they must
individually vanish. In particular this implies that $M=0$ and
$F^{+}_{\mu \nu }=0$. This is a `vanishing' theorem.

Even when the scalar curvature is not positive the squaring argument puts
strong constraints on the solution set. As
\be
\int_{X}d^{4}x \sqrt{g}\frac{1}{2}\left( |M|^{2} + \frac{1}{4}R \right)^{2}
\geq 0 \,
\ee
we conclude that
\be
\int_{X}d^{4}x \sqrt{g}\left( \frac{1}{2}|M|^{4} + \frac{1}{4}R|M|^{2} \right)
\geq -\frac{1}{32}\int_{X}d^{4}x \sqrt{g} R^{2} \, .
\ee
Now, re-write (\ref{blw}) as
\be
\int_{X}d^{4}x\sqrt{g}\left( \frac{1}{2}|F^{+}|^{2} +
 |DM|^{2} \right) = - \int_{X}d^{4}x \sqrt{g} \left(
\frac{1}{2} |M|^{4}+ \frac{1}{4}R|M|^{2} \right) \, ,
\ee
which yields an inequality
\be
\int_{X}d^{4}x\sqrt{g}|F^{+}|^{2}  \leq - \int_{X}d^{4}x \sqrt{g} \left(
\frac{1}{2} |M|^{4} +\frac{1}{4}R|M|^{2} \right) \, .
\ee
Combining the two inequalities yields
\be
\int_{X}d^{4}x\sqrt{g} \frac{1}{2}|F^{+}|^{2} \leq  \frac{1}{16}
 \int_{X}d^{4}x \sqrt{g} R^{2} \, .
\ee

The line bundle in question, $L$, satisfies $c_{1}(L)^{2}=(2\chi(X) +
3\tau(X))/4$, but we can also express this as
\be
c_{1}(L)^{2}= \frac{1}{(2\pi)^{2}}\int_{X}F^{2}=
\frac{1}{(2\pi)^{2}}\int_{X}\sqrt{g}\left(|F^{+}|^{2}-|F^{-}|^{2}\right)
{}.
\ee
When the dimension of the moduli space vanishes the bound on $F^{+}$
also places a bound on $F^{-}$,
\be
\int_{X}d^{4}x \sqrt{g}|F^{-}|^{2} \leq \frac{1}{16}\int_{X}d^{4}x
\sqrt{g}R^{2}  - \pi^{2} (2\chi(X) + 3\tau(X) ) .
\ee
This places a bound on the number of $x$'s that will lead to a zero
dimensional moduli space. Hence, associated to every four manifold, there
will only be a finite number of invariants $n_{x}$.
One can also read the inequality as a condition on the underlying four
manifold, namely there will be a zero dimensional moduli space only if
\be
\frac{1}{(4\pi)^{2}}\int_{X}d^{4}\sqrt{g}R^{2} \geq 2\chi(X) + 3\tau(X) .
\ee

\subsection{K\"{a}hler Manifolds}
When $X$ is a K\"{a}hler manifold we may decompose the components of $M$
according to

\vspace{.5cm}
\noindent \underline{$\Omega_{+}= \o.\Omega^{0}\oplus \Omega^{(2,0)}\oplus
\Omega^{(0,2)}$}

Let us choose our complex co-ordinates on $\RR^{4}$ to be
$z^{1}=x^{1}+ix^{2}$, $z^{2}=x^{3}+ix^{4}$. The $(2,0)$ and the $(0,2)$
forms are spanned by
\bea
dz^{1}dz^{2}& = & \left(dx^{1}dx^{3}- dx^{2}dx^{4}\right) +i \left(
dx^{1}dx^{4}+dx^{2}dx^{3} \right) \nonumber \\
d\zb^{1}d\zb^{2} &=& \left(dx^{1}dx^{3}- dx^{2}dx^{4}\right) -i \left(
dx^{1}dx^{4}+dx^{2}dx^{3} \right)
\eea
The symplectic two form, $\o$ can be taken to be
\be
\o = \frac{i}{2}dz^{1}d\zb^{1} + \frac{i}{2}dz^{2}d\zb^{2}=
dx^{1}dx^{2}+ dx^{3}dx^{4} \, .
\ee
Self dual two forms $\Phi$ satisfy
\be
\Phi_{\a \b}= \frac{1}{2}\eps_{\a \b \mu \nu} \Phi^{\mu \nu},
\ee
so that
\bea
\Phi &=& 2 \Phi_{12}\left(dx^{1}dx^{2}+dx^{3}dx^{4}\right)
+ 2 \Phi_{13}\left(dx^{1}dx^{3}-dx^{2}dx^{4}\right) +
2 \Phi_{14}\left(dx^{1}dx^{4}-dx^{2}dx^{3}\right) \nonumber \\
&=& 2 \Phi_{12}\o + \left(\Phi_{13}-i\Phi_{14}\right) dz^{1}dz^{2}+
\left(\Phi_{13}+i\Phi_{14}\right) d\zb^{1}d\zb^{2} \, .
\eea

{}From this we see that we may decompose the space of self dual two forms,
$\Omega_{+}$, as $\o \Omega^{(0,0)} \oplus \Omega^{(2,0)} \oplus
\Omega^{(0,2)}$. This decomposition holds on any K\"{a}hler surface.

\vspace{1cm}

\noindent \underline{$(K^{1/2}\otimes L) \oplus (K^{-1/2}\otimes L)$}

Not only does one have a nice decomposition of two forms on a K\"{a}hler
surface but there is also a nice decomposition of the charged spinors
when $X$ is spin. Indeed $S^{+} \otimes L \cong (K^{1/2}\otimes L)
\oplus (K^{-1/2} \otimes L )$, where $K$ is the canonical
bundle\footnote{The canonical line bundle is made up of top
degree holomorphic differentials.} and $K^{1/2}$ is some square root.
When $X$ is not spin, then $(K^{1/2}\otimes L)$ is well defined as a
square root of the line bundle $(K \otimes L^{2})$.

The upshot of this discussion is that we may decompose the spinor $M$
into two components $\a \in (K^{1/2} \otimes L)$ and $-i \overline{\b}
\in (K^{-1/2} \otimes L$.

%\begin{center}
%\fbox{To be continued}
%\end{center}

\vspace{1cm}
\noindent \underline{Computations}

After these preliminaries we find that the `monopole' equations take on
the following simple form
\bea
F^{(2,0)}&=& \a \b  \nonumber \\
F^{(1,1)}_{\o} &=& -\frac{\o}{2}\left( |\a|^{2}-|\b|^{2} \right)
\nonumber \\
F^{(0,2)}&=&
 \overline{\a}\overline{\b} \, . \label{ksw}
\eea

In this notation (\ref{blw}) can be rewritten
\bea
\int_{X}d^{4}x \sqrt{g}\left(\frac{1}{2}|s|^{2} + |k|^{2} \right) &=&
\int_{X}d^{4}x \sqrt{g}\left( \frac{1}{2}|F^{+}|^{2} + g^{\mu \nu}
\overline{D_{\mu}\a}D_{\nu}\a + g^{\mu \nu} \overline{D_{\mu}}\b
D_{\nu}\overline{\b} \right. \nonumber \\
& & \; \; \; \; \left. \frac{1}{2}(|\a|^{2}+ |\b|^{2})^{2}+\frac{1}{4}R(
|\a|^{2}+ |\b|^{2}) \right) \, . \label{kblw}
\eea

We notice that the right hand side has a symmetry
\be
A \rightarrow A \, , \; \; \; \a \rightarrow \a \, , \; \; \; \b
\rightarrow - \b \, .
\ee
Even though this is not a symmetry of the equations, it does have strong
implications. Firstly, the right hand side of (\ref{kblw}) has a zero
only at a solution of the monopole equations (as from the left hand side
we would require $s=k=0$). If for some $\b$ the right hand side is zero
then, by the symmetry, it must be also zero for $-\b$. Thus if
$(A,\a,\b)$ is a solution to the monopole equation so is $(A,\a,-\b)$.

The situation just described implies that
\be
F^{(2,0)}= \a \b = -\a \b \, .
\ee
We have thus learnt that
\be
0=F^{(2,0)}=F^{(0,2)}=\a \b = \overline{\a}\overline{\b} \, . \label{hol}
\ee
The vanishing of the $(0,2)$ component of the curvature tells us that
the line bundle $L$ is a holomorphic line. The basic classes $x$ are
therefore of type $(1,1)$ for any K\"{a}hler manifold, which is quite
restrictive.
The equations (\ref{hol}) imply that either $\a=0$ or $\b=0$. We can
deduce which
of the two is zero by integrating the $(1,1)$ part of (\ref{ksw}),
\be
\frac{1}{2\pi}\int_{X}\o F = -\frac{1}{4\pi}\int_{X}\o \o
\left(|\a|^{2}-|\b|^{2}\right) \, .
\ee
The left hand side is the degree of the holomorphic line bundle $L$,
sometimes denoted by $\deg(L)$, which is a topological
invariant. When $\deg(L)=0$, there is the possibility of having trivial
instantons (in this case both $\a$ and $\b$ must be zero) and we
consider metrics for which this is not possible. Now if $\deg(L) > 0$ we
must have $\a \neq 0$ and $\b =0$. Hence, the topological data all but
fixes the solutions! I will not describe the geometric meaning of the
$(1,1)$ part of the equations (\ref{ksw}) but refer the reader to
\cite{witsw}.

The equations for the spinors are
\bea
\overline{\partial}_{A}\a -i \overline{\partial}_{A}^{*} \overline{\b} & = &
0  \, , \nonumber
\\
\overline{\partial}_{A} \b + i\overline{\partial_{A}}^{*}\overline{\a} &=& 0
 \,
\eea
but we will not have need of these.
\subsection{Perturbation}
For a K\"{a}hler manifold the condition $b_{2}^{+}>1$ is equivalent to
$H^{(2,0)}(X) \neq 0$. In this case we can take $p = \eta
+\overline{\eta}$, where $\eta$ is a non-zero holomorphic two-form.
Before perturbing, the first Chern class of the line bundle $L$ was
given completely by the $(1,1)$ component of $F$. The perturbation is
chosen so that this remains the case, namely
\be
\int_{X}F^{(2,0)}\overline{\eta}=\int_{X}F^{(0,2)}\eta =0 \, .
\ee
An argument similar to the one that lead to (\ref{hol}) yields
\be
0=F^{(2,0)}= \a \b - \eta \, .
\ee

The vanishing of $F^{(0,2)}$ means that we are still in the realm of
holomorphic bundles. The important equation is
\be
\a \b = \eta \, . \label{holdiv}
\ee
Now $\eta \in H^{(2,0)}(X)$ $ (\equiv H^{0}(X,K))$, and $\a$ and $\b$
are holomorphic sections of $K^{1/2}\otimes L^{\pm 1}$. Let the divisor
of $\eta$ be a union of irreducible components $\S_{i}$ with
multiplicity $r_{i}$. Then
\be
c_{1}(K) = \sum_{i}r_{i}[\S_{i}] \, ,
\ee
where $[\S_{i}]$ denotes the cohomology class that is Poincar\'{e} dual to
the Riemann surface $\S_{i}$. As always we take the $\S_{i}$ to span
$H_{2}(X,\ZZ)$, and consequently $[\S_{i}]$ to span $H^{2}(X,\ZZ)$. The
integers $r_{i} \geq 0$ as the sections are holomorphic. Likewise as
$\a$ is a holomorphic section of $K^{1/2}\otimes L$ and $\b$ a
holomorphic section of $K^{1/2}\otimes L^{-1}$
\bea
c_{1}(K^{1/2} \otimes L) &=& \sum_{i} s_{i}[\S_{i}] \nonumber \\
c_{1}(K^{1/2} \otimes L^{-1}) &=& \sum_{i} t_{i}[\S_{i}] \, ,
\eea
with $0 \leq s_{i}$, and $0 \leq t_{i}$. For line
bundles, $E$ and $F$, $c_{1}(E \otimes F) = c_{1}(E) + c_{1}(F)$. Set
\be
c_{1}(L)= \sum_{i} u_{i}[\S_{i}] ,
\ee
where there is no a priori constraint on the sign of the integers
$u_{i}$. Even though we do not know the sign of the $u_{i}$ we do know
that
\be
t_{i}= \frac{1}{2}r_{i} - u_{i} \geq 0
\ee
so that $r_{i} \geq 2u_{i}$. We also know that
\be
s_{i} = \frac{1}{2}r_{i} + u_{i}
\ee
but, because of the bound on the $u_{i}$, we obtain $0 \leq s_{i} \leq
r_{i}$. As we could have run through the argument with $s_{i}$ and
$t_{i}$ interchanged we conclude that $0 \leq t_{i} \leq r_{i}$.

%Because of the factorisation
%(\ref{holdiv}), one has
%\be
%2r_{i}=s_{i} + t_{i} \, ,
%\ee
%so that $0 \leq s_{i} \leq 2r_{i}$. The first Chern class of $L$, is
%\bea
%c_{1}(L) &=& c_{1}(K^{1/2}\otimes L) - c_{1}(K^{1/2}) \nonumber \\
%         &=& \sum_{i}(s_{i}-\frac{1}{2}r_{i})[\S_{i}] \, .
%\eea
The basic class is of the form $x=-2c_{1}(L)$ or
\be
x= - \sum_{i}(2s_{i}-r_{i})[\S_{i}] \, . \label{x}
\ee

With the perturbed moduli space having dimension zero the invariants
associated with monopole moduli spaces of higher dimension vanish.
Finally the basic classes $x$ are of the form (\ref{x}) and satisfy
(\ref{dim0}), i.e. $x^{2} = c_{1}(K)^{2}$. Notice that on any K\"{a}hler
surface there are always the classes $x = \pm c_{1}(K)$ which happens
only when $s_{i}=0$ or $s_{i = r_{i}}$, respectively, and so these yield
$n_{x} = \pm 1$.

The operator that determines the relative signs of the contributions is
$T$ that appeared in the linearisation of the monopole equations
(\ref{Top}). A consistent choice of sign comes by giving the sign of
$\det {T}$ for each $x$; just as one needs to give the sign of $\det{
H_{P} (f)}$ for the Euler character at each $P$. Unfortunately, explaining
how to specify the sign
of $\det{T}$ would take us too far afield and I refer the reader to
\cite{witsw}.

%\begin{center}
%\fbox{To be continued}
%\end{center}
\subsection{Implications}
After all this work, we are in a position to make use of these
invariants to learn something new about four-manifolds.

Using the fact that on K\"{a}hler manifolds the Seiberg-Witten
invariants are {\bf not} zero one can prove that algebraic surfaces do not
have decompositions into connected sums\footnote{The
connected sum $X \# Y$ of two manifolds is obtained on removing a four
disc from both $X$ and $Y$ and then glueing them together along their
boundary $S^{3}$.} with $b_{2}^{+} >0$ on both sides. One
proves this by showing that the Seiberg-Witten invariant {\bf must} be zero
for a connected sum $ X \# Y$ which each have $b_{2}^{+}>0$. This is the
same line of argument used in Donaldson theory to prove the same result.
The simplification here is that we are dealing with $U(1)$ and not
$SO(3)$.

Consider a
metric on $X\# Y$ in which $X$ and $Y$ are connected by a long neck of
the form $S^{3} \times I$, with $I$ an interval of $\RR$. As we stretch
the neck out, and make it longer and longer, any solution of the monopole
equations will vanish in the neck (by the vanishing theorem) since $R > 0$ on
$S^{3}$. Now one may define a $U(1)$ action on ${\cal M}$ by gauge
transforming the solutions on $Y$ by a constant gauge transformation,
leaving the fields on $X$ fixed. A fixed point of this $U(1)$ action
would be a solution for which $M$ vanishes on $X$ or on $Y$. But for
generic metrics on both sides there is no such solution (except with
$A=0$). Now the dimension of ${\cal M}$ is zero and
there is no free $U(1)$ action on a set of points unless it is the empty
set.  Consequently the invariants for such connected sums must be zero,
which is a contradiction.

Since Wittens paper appeared there has been much activity in the
mathematics community. An old conjecture of Thom has been proved using
these invariants \cite{km2}. One should watch the xxx archives for
more results!

\appendix

\section{ Conventions}
Gamma matrices $(\ga^{a})^{\b}_{\; \a}$ are taken to be hermitian and to
satisfy
\be
\{ \ga^{a} , \ga^{b} \} = 2 \d^{ab} .
\ee
The definition and properties of $\ga_{5}$ are
\be
\ga_{5} = - \prod_{1}^{4} \ga^{a} \, , \; \; \ga_{5}^{2} =1 \, , \; \;
\ga_{5}^{\dagger} = \ga_{5} \, .
\ee
The matrices $\sigma_{ab}$ are given by
\be
\sigma_{ab} = \frac{1}{2} [\ga_{a},\ga_{b}] \, ,
\ee
with the flat labels $\{ a \}$ raised and lowered with the kronecker
$\d_{ab}$. The following is now easy to check
\be
\sigma_{ab}\ga_{5} =  \frac{1}{2} \eps_{abcd} \sigma^{cd}
\ee
with $\eps_{1234}=1$.

Spinors carry their labels `upstairs' i.e. $M^{\a}$. A positive
chirality spinor (one writes $M \in S_{+}$) satisfies
\be
\ga_{5 \; \b}^{\, \a}M^{\b} = M^{\a} .
\ee
The hermitian conjugate of a spinor is $\overline{(M^{\a})} =
\overline{M}_{\a}$. In particular if $M \in S_{+}$ then as $\ga_{5}$ is
hermitian, we have
\be
\overline{M}_{\b}\ga_{5 \; \a}^{ \, \b} = \overline{M}_{\a} .
\ee
One writes $\overline{M} \in S_{-}$.

\section{ Vierbeins and the Spin Connection}

We will recall here some of the basic formula for the coupling of spinors
to a gravitational field. One introduces a vierbein $e_{\mu}^{a}$, where
the label $a$ is an internal Lorentz label. The gauge field for the
internal gauge group $SO(4)$ is called the spin connection and is denoted
by $\o_{\mu}^{ab}$.

The covariant derivative is
\be
D_{\mu} M = \left(\partial_{\mu} + \frac{1}{4} \o_{\mu}^{ab}\sigma_{ab}
\right) M \, .
\ee
The matrices $\sigma_{ab} = \frac{1}{2}[\ga_{a},\ga_{b}]$, satisfy the
algebra of the Lorentz group, namely
\be
[\sigma_{ab},\sigma_{cd}] = \d_{ad} \sigma_{bc}-\d_{bd}\sigma_{ac} +
\d_{bc}\sigma_{ad}-\d_{ac}\sigma_{bd} \, .
\ee
The spin connection is determined from the fact that one
requires that the covariant derivative of the vierbein vanishes
\be
D_{\mu}e_{\nu}^{a}= \partial_{\mu} e^{a}_{\nu} - \Gamma_{\mu \nu}^{\lambda}
e^{a}_{\lambda} +
\o_{\mu b}^{a} e^{b}_{\nu} = 0 \,
\ee

With this one calculates that
\bea
[D_{\mu},D_{\nu}] &=& \left( \partial_{\mu}\o_{\nu}^{ab}-
\partial_{\nu}\o_{\mu}^{ab} +[\o_{\mu},\o_{\nu}]^{ab}\right)
\frac{1}{4}\sigma_{ab} \nonumber \\
 &=& R_{\mu \nu}^{ab} \frac{1}{4}\sigma_{ab} \, .
\eea
This gives back the definition of the Riemann curvature tensor

A property of the Riemann curvature tensor that will be useful is
\be
R_{\kappa \lambda \mu \nu}+ R_{\nu \lambda \kappa \mu }
+ R_{ \mu \lambda \nu \kappa } =0 \, . \label{cyclic}
\ee

Consider
\bea
\frac{1}{2}\ga^{\mu}\ga^{\nu}[D_{\mu},D_{\nu}] &=& \frac{1}{8}  \left(
\partial_{\mu} \o_{\nu} -\partial_{\nu}\o_{\mu} + [\o_{\mu},\o_{\nu}]
\right)^{ab}  \ga^{\mu}\ga^{\nu}\sigma_{ab} \nonumber \\
& = &  \frac{1}{8}\left(
\partial_{\mu} \o_{\nu} -\partial_{\nu}\o_{\mu} + [\o_{\mu},\o_{\nu}]
\right)_{\kappa \lambda }  \ga^{\mu}\ga^{\nu} \ga^{\kappa}
\ga^{\lambda} \nonumber \\
& =&  \frac{1}{8} R_{\mu \nu \kappa \lambda} \ga^{\mu}\ga^{\nu}
\ga^{\kappa} \ga^{\lambda}
\eea

We can use the identity (\ref{cyclic})
\bea
0 &= & \left( R_{\kappa \lambda \mu \nu}
+ R_{\nu \lambda \kappa \mu } + R_{ \mu \lambda\nu \kappa}
\right)\ga^{\mu}\ga^{\nu}\ga^{\kappa}
\ga^{\lambda} \nonumber \\
& =& R_{\kappa \lambda \mu \nu} [\ga^{\mu}\ga^{\nu}\ga^{\kappa}
 + \ga^{\kappa}\ga^{\mu}\ga^{\nu}
+\ga^{\nu}\ga^{\kappa} \ga^{\mu}]\ga^{\lambda} \, ,
\eea
and standard $\ga$ matrix-ology, such as
\be
\ga^{\nu}\ga^{\kappa}\ga^{\mu} = 2g^{\kappa \mu}\ga^{\nu}-2g^{\nu \mu}
\ga^{\kappa} + \ga^{\mu}\ga^{\nu}\ga^{\kappa}
\ee
to write all the products of three gamma matrices in the order $
\ga^{\mu}\ga^{\nu}\ga^{\kappa}$. One may now deduce that
\be
 R_{\kappa \lambda \mu \nu}  \ga^{\mu}\ga^{\nu}\ga^{\kappa} \ga^{\lambda}
= -2 g^{\kappa \mu} g^{\lambda \nu} R_{\kappa \lambda \mu \nu} = -2R \, .
\ee

%\subsection*{Acknowledgements}

\rnc{\Large}{\normalsize}


\begin{thebibliography}{99}
\addcontentsline{toc}{section}{References}
\frenchspacing
\small
\addtolength{\itemsep}{-4pt}
\bibitem{ak} S. Akbulut, {\bf Lectures on Seiberg-Witten Invariants},
           alg-geom/9510012, (1995).
\bibitem{atJef}  M. Atiyah and L. Jeffrey, {\bf Topological Lagrangians and
                Cohomology}, J. Geom. Phys. 7 (1990) 120.
%\bibitem{aps}  M. Atiyah and I. Singer, {\bf ..}, Math. Proc.
%               Camb. Phil. Soc. 77 (1975) 43.
\bibitem{brt}   D. Birmingham, M. Rakowski and G. Thompson, {\bf
                Topological Field Theories, Nicolai Maps and BRST
                Quantization},  Phys. Lett. B212 (1988) 187.
\bibitem{bbrt}   D. Birmingham, M. Blau, M. Rakowski and G. Thompson, {\bf
	       Topological Field Theory}, Phys. Rep. 209 4 \& 5 (1991).
\bibitem{blau}  M. Blau, {\bf The Matthai-Quillen formalism and Topological
                Field Theory}, J. Geom. Phys. 11 (1993) 95.
\bibitem{bjt} M. Blau, I. Jermyn and G. Thompson, {\bf Solutions to
              certain topological field theories on particular three
              manifolds}, in preparation.
\bibitem{bt}  M. Blau and G. Thompson, {\bf Topological gauge theories
             of antisymmetric tensor fields}, Ann. Phys. 205 (1991) 130.
\bibitem{btym}  M. Blau and G. Thompson, {\bf Quantum Yang-Mills theory
              on arbiitrary surfaces}, Int.
               J. Mod. Phys. A7 (1992) 3781.
\bibitem{btver} M. Blau and G. Thompson, {\bf Derivation of the Verlinde
		 formula from Chern-Simons theory and the $G/G$ model},
		 Nucl. Phys. B408 (1993)345.
\bibitem{btrev} M. Blau and G. Thompson, {\bf Lectures on 2d Gauge
               Theories}, ICTP series vol. 10, eds. E. Gava et. al., World
             Scientific, (1994).
\bibitem{btcasson} M. Blau and G. Thompson, {\bf  $N=2$ Topological Gauge
               Theories, the Euler Characteristic of Moduli spaces and the
             Casson Invariant}, Commun. Math. Phys. 152 (1993) 41.
\bibitem{braam} P. Braam, {\bf Monopoles and 4-Manifolds}, Lectures in the
conference on topological and geometrical problems related to quantum field
theory, Trieste March 1995, unpublished.
\bibitem{ruibin} A. Carey, J. McCarthy, B. Wang and R. Zhang, {\bf
               Topological quantum field theory and Seiberg-Witten
              monopoles},  hep-th 9504005.
\bibitem{cmr} S. Cordes, G. Moore and S. Ramgoolam, {\bf Lectures on 2D
            Yang Mills theory, equivariant cohomology and topological string
            theory}, 1994 Trieste Spring School; hep-th/9411210.
\bibitem{dijk} R. Dijkgraaf, {\bf 4-Manifolds and Gauge Theories}, 1995
               Spring School Trieste, unpublished.
\bibitem{dk} S. Donaldson and P. Kronheimer, {\bf The geometry of
four-manifolds}, Oxford Science Publications, (1990).
\bibitem{egh}      T. Eguchi, P. Gilkey and A. Hanson, {\bf Gravitation,
               Gauge Theories and Differential Geometry}, Phys. Rep. 66
              (1980) 213-393.
\bibitem{dfine} D. Fine, {\bf Quantum Yang-Mills on a Riemann Surface},
              Commun. Math. Phys. 140 (1991) 321.
\bibitem{fu} D. Freed and K. Uhlenbeck, {\bf Instantons and
four-manifolds}, MSRI publications, Springer (1984).
\bibitem{gsw} M. Green, J. Schwarz and E. Witten, {\bf Superstring
              Theory: Vol 2}, Cambridge Univ. Press (1987).
\bibitem{hp}  S. Hawking and C. Pope, {\bf Generalised Spin Structures
              in Quantum Gravity}, Phys. Lett B73 (1978) 42.
\bibitem{km}  P. Kronheimer and T. Mrowka, {\bf Recurrence Relations and
              Asymptotics for Four-Manifold Invariants}, Bull. Amer.
             Math. Soc. 30 (1994) 215.
\bibitem{km2} P. Kronheimer and T. Mrowka, {\bf The genus of embedded
             surfaces in the projective plane}, Math. Research Letters 1
            (1994) 797.
\bibitem{lm} H. Lawson and M. Michelsohn, {\bf Spin Geometry}, Princeton,
            New Jersey, (1989).
\bibitem{l}  A. Lichnerowicz, {\bf Spineurs Harmoniques}, C.R. Acad.
             Sci. Ser. A257 (1963) 7.
\bibitem{mansfield} P. Mansfield, {\bf The first Donaldson Invariant as the
                 Winding Number of a Nicolai Map}, Mod. Phys. Lett. A3
               (1988) 1647.
\bibitem{marc} M. Marcolli, {\bf Notes on Seiberg-Witten Gauge Theory},
               dg-ga/9509005, (1995).
\bibitem{MatQuil} V. Matthai and D. Quillen, {\bf Superconnections, Thom
                Classes and Equivariant Differential Forms}, Topology
                25 (1986) 85.
\bibitem{rusakov} B. Rusakov, {\bf Loop averages and partition function
                  in $U(N)$ gauge theory on two dimensional manifolds },
                  Mod. Phys. Lett. A5
                (1990) 693.
\bibitem{sw} N. Seiberg and E. Witten, {\bf Electric- magnetic duality,
monopole condensation and confinement in $N=2$ Yang-Mills theory}, Nucl.
Phys. B246 (1994) 19.
\bibitem{thompson} G. Thompson, {\bf 1992 Trieste Lectures on
                  topological gauge theory and Yang-Mills theory}, ICTP
series vol 9, eds: E. Gava et al, World Scientific, (1993).
\bibitem{taubes1} C. Taubes, {\bf The Seiberg-Wittem invariants and
symplectic forms}, Math. Research Letters, 1 (1994) 809.
\bibitem{vafwit} C. Vafa and E. Witten, {\bf A Strong Coupling Test of
               S-Duality}, hep-th/9408074.
\bibitem{ver} E. Verlinde, {\bf Fusion rules and modular transformations
               in 2d conformal field theory}Nucl. Phys. B300 (1988) 360;
\bibitem{versdual} E. Verlinde, {\bf Global Aspects of Electric-Magnetic
                Duality}, hep-th/9505.
\bibitem{witmorse} E. Witten, {\bf Supersymmetry and Morse Theory}, J. Diff.
                  Geom. 17 (1982) 661.
\bibitem{wit2ddon} E. Witten, {\bf Two dimensional gauge theories
revisited}, J. Geom. Phys. 9 (1992) 303.
\bibitem{witsdual} E. Witten, {\bf On S Duality in Abelian Gauge
                 Theory}, hep-th/9505186.
\bibitem{witsw} E. Witten, {\bf Monopoles and Four Manifolds}, Math. Res.
               Lett. 1 (1994) 769.
\bibitem{witsym} E. Witten, {\bf Supersymmetric Yang-Mills Theory on a
               Four-Manifold}, J. Math. Phys. 35 (1994) 5101.


%	      R. Dijkgraaf and E. Verlinde, Nucl. Phys. B (Proc. Suppl.)
%	      5B (1988) 87.
\bibitem{Witcs} E. Witten, {\bf Quantum field theory and the Jones
              polynomial}, Commun. Math. Phys. 121 (1989) 351.
%\bibitem{ms} G. Moore and N. Seiberg, Commun. Math. Phys. 123 (1989) 177.
%\bibitem{nnver} R. Bott, Int. J. Mod. Phys. A6 (1991) 2847; A. Szenes,
%           Duke Math. Journal 64 (1991) R93; A. Bertram and
%           A. Szenes, {\em Hilbert polynomials of moduli spaces of rank $2$
%           vector bundles II}, Harvard preprint (1991); L. Jeffrey and
%           J. Weitsman, Commun. Math. Phys. 150 (1992) 593;
%           G. Daskalopoulos and R. Wentworth, Commun. Math. Phys. 151 (1993)
%           437; B. van Geemen and E. Previato, {\em Prym
%           varieties and the Verlinde formula}, MSRI Berkeley preprint (1991);
%           M. Thaddeus, {\em Stable pairs, linear systems and the Verlinde
%           formula}, MSRI Berkeley preprint (1992); M.S. Narasimhan and T.R.
%           Ramadas, {\em Factorisation of generalized theta functions I}, ICTP
%           Trieste preprint (1993).
%\bibitem{sp} M. Spiegelglas, quoted in \cite{ewwzw}; M. Spiegelglas and S.
%	     Yankielowicz, {\em G/G topological field theories by cosetting
%	     $G_{k}$}, Nucl. Phys. B393 (1993)
%\bibitem{verver} E. Verlinde and H. Verlinde, {\em Conformal field theory
%		 and geometric quantization}, in {\em Superstrings '$89$}
%		 (eds. M. Green et al.), World Scientific, Singapore (1990)
%		 422.
%\bibitem{ewwzw} E. Witten, Commun. Math. Phys. 144 (1992) 189.
%\bibitem{ewwz} E. Witten, Non-Abelian Bosonization
%\bibitem{wz} Wess-Zumino
%\bibitem{pw} Polyakov-Wiegmann
%\bibitem{gko} GKO
%\bibitem{yan} Yankielowicz et al.
%\bibitem{csknots} some ref on CS and knot theory
%\bibitem{gross} ym, strings, and wilson loops
%\bibitem{kostov} russians on wilson loops
%\bibitem{botu} Bott and Tu
%\bibitem{btbf2} M. Blau and G. Thompson, Ann. Phys. 205 (1991) 130.
%\bibitem{btym}  M. Blau and G. Thompson, Int. J. Mod. Phys. A7 (1992) 3781.
%\bibitem{ewym}  E. Witten, Commun. Math. Phys. 141 (1991) 153.
%\bibitem{ew2d}  E. Witten, J. Geom. Phys. 9 (1992) 303.
%\bibitem{emss}   S. Elitzur, G. Moore, A. Schwimmer and N. Seiberg, Nucl.
%		 Phys. B326 (1989) 108.
%\bibitem{axcs}   S. Axelrod, S. della Pietra and E. Witten, J. Diff. Geom.
%		 33 (1991) 787; S. Axelrod, {\em Geometric Quantization of
%		 Chern-Simons Gauge Theory}, Ph. D. thesis, Princeton
%		 University, (1991).
%\bibitem{rs}   D. Ray and I. Singer, Adv. Math. 7 (1971) 145; {\em
%               Analytic torsion}, in {\em
%               Partial Differential Equations}, Proc. Symp. Pure Math.
%               23 (1973) 167.
%\bibitem{nncs} D. Freed and R. Gompf, Commun. Math. Phys. 141 (1991) 79;
%	       L. Jeffrey, Commun. Math. Phys. 147 (1992) 563;
%	       L. Rozansky, {\em A large $k$ asymptotics of Witten's
%	       invariant of Seifert manifolds}, Texas preprint UTTG-06-93.
%\bibitem{qdet} D. Quillen, Funt. Anal. Appl. 19 (1986) 31.
%\bibitem{asrs} A. Schwarz, Lett. Math. Phys. 2 (1978) 247.
%\bibitem{dfcs} D. Freed, J. reine angew. Math. 429 (1992) 75.
%\bibitem{hobf} G. Horowitz, Commun. Math. Phys. 125 (1989) 417.
%\bibitem{btbf1}  M. Blau and G. Thompson, Phys. Lett. B228 (1989) 352.
%\bibitem{gk}  K. Gawedzki and A. Kupiainen, Nucl. Phys. B320 (1989) 625.
%\bibitem{bgv} N. Berline, E. Getzler and M. Vergne, {\em Heat Kernels and
%	      Dirac Operators}, Springer, New York (1991).
%\bibitem{gacs}  K. Gawedzki, {\em Wess-Zumino-Witten conformal field theory},
%	      in {\em Constructive Quantum Field Theory II} (eds. G. Velo
%	      and A. Wightman), Plenum, New York (1990) 89.
%\bibitem{gi} P. Gilkey, {\em Invariance Theory, the Heat Equation, and the
%	     Atiyah-Singer Index Theorem}, Publish or Perish (1984).
%\bibitem{imcs} C. Imbimbo, Phys. Lett. B258 (1991) 353.
%\bibitem{btd} T. Br\"{o}cker and T. tom Dieck, {\em Representations of Compact
%	    Lie Groups}, Springer, New York (1985).
\end{thebibliography}
\end{document}